\documentclass[english,prb]{revtex4}
\usepackage[T1]{fontenc}
\usepackage[latin9]{inputenc}
\usepackage{bm}
\usepackage{amsmath}
\usepackage{graphicx}
\usepackage{amssymb}

\makeatletter

\@ifundefined{textcolor}{}
{%
 \definecolor{BLACK}{gray}{0}
 \definecolor{WHITE}{gray}{1}
 \definecolor{RED}{rgb}{1,0,0}
 \definecolor{GREEN}{rgb}{0,1,0}
 \definecolor{BLUE}{rgb}{0,0,1}
 \definecolor{CYAN}{cmyk}{1,0,0,0}
 \definecolor{MAGENTA}{cmyk}{0,1,0,0}
 \definecolor{YELLOW}{cmyk}{0,0,1,0}
 }

\makeatother

\usepackage{babel}

\begin{document}

\title{Charge instabilities and topological phases in the  extended Hubbard model on the honeycomb lattice with enlarged unit cell}
\author{Adolfo G. Grushin$^{1}$}
\author{ Eduardo V. Castro$^{2}$}
\author{Alberto Cortijo$^{1}$}
\author{Fernando de Juan$^{3,4}$}
\author{Mar\'{\i}a A. H. Vozmediano$^{1}$}
\author{Bel\'en Valenzuela $^{1}$}

\affiliation{$^{1}$Instituto de Ciencia de Materiales de Madrid, CSIC, Cantoblanco,
E-28049 Madrid, Spain}

\affiliation{$^{2}$CFIF, Instituto Superior 
T\'ecnico, TU Lisbon, Av. Rovisco Pais, 1049-001 Lisboa, Portugal}

\affiliation{$^{3}$Lawrence Berkeley National Laboratory,
1 Cyclotron Rd, Berkeley, CA 94720}

\affiliation{$^{4}$Department of Physics, University of California, Berkeley, CA 94720, USA}

\date{\today}

\begin{abstract}
We study spontaneous symmetry breaking in a system of spinless fermions in the Honeycomb lattice paying special emphasis to the role of an enlarged unit cell on time reversal symmetry broken phases. We use a tight binding model with nearest neighbor hopping $t$ and Hubbard interaction $V_1$ and $V_2$ and extract the phase diagram as a function of electron density and interaction within a mean field variational approach. The analysis completes the previous work done in Phys. Rev. Lett. 107, 106402 (2011) where phases with non--trivial topological properties were found with only a nearest neighbor interaction $V_1$ in the absence of charge decouplings. We see that the topological phases are suppressed by the presence of metallic charge density fluctuations. The addition of  next to nearest neighbor interaction $V_2$ restores the topological non-trivial phases. 
\end{abstract}
\maketitle
\section{Introduction}

\label{sec:intro}

Topological phases of matter are a new paradigm in condensed matter
physics.\cite{H04,H88,Vo03,QFTmbs} These phases evade a standard classification
in terms of local order parameters and broken symmetries, being 
described by topological invariants. In addition to the
obvious interest from a fundamental viewpoint, the robustness of topological
properties against certain local perturbations make these phases appealing
also in applied physics. The recently discovered time reversal invariant topological insulators
are a raising star in the family, with topological invariants protected
against non magnetic disorder by time reversal symmetry ($\mathcal{T}$).\cite{BHZ06,HK10,QZ11}
Together with time reversal invariant topological insulators, the quantum Hall insulating state is the paradigmatic example of a topological phase.\cite{E08} The non trivial topology is in this case driven by an external magnetic field which breaks $\mathcal{T}$. In quantum anomalous Hall (QAH) insulators, another example of a topological phase, $\mathcal{T}$ is broken spontaneously, and a quantized anomalous Hall (AH) conductivity arises in the absence of any external magnetic field.\cite{NSetal10} When the Fermi level does not fall into the bulk band gap there is a non quantized contribution to the AH conductivity  characterized as a property of the Fermi surface  through its Berry phase and the systems are termed topological Fermi liquids. \cite{H04,SF08}

The Honeycomb lattice is perhaps the best studied case for its special properties. \cite{CGV10,CGV12} It is bipartite and yet due to its  topology it is easily amenable to frustration. It is fair to say that the pioneer works proposing this lattice  as a ``condensed matter simulation of a three-dimensional anomaly"  \cite{S84,H88}, together with the analysis of the spin orbit in graphene done in  refs. \onlinecite{KM05,KM05b} opened the modern field of topological insulators.  In these pioneering works the breaking of time reversal symmetry that allowed the topological non--trivial phases was explicitly put in the Hamiltonian in the form of complex hopping parameters. The non interacting behavior of the topological phases in the insulating family
is at present fairly well understood, and the attention is now
shifting to the effect of electron-electron interactions on these
phases\cite{PB10,VSetal11,Rachel10,SR10,HLA11,ZWZ10,WSZ+10,hailee11,ACS12}
and to the nature of the phase transitions between topological and ordinary 
phases (see \cite{VSetal11} and references therein). 
In most cases following the original work by Haldane \cite{H88} breaking $\mathcal{T}$ is associated with \emph{bond} order (complex hoppings) originating finite flux states.\cite{RQetal08,SYetal09,LYM10,WRetal10,WF10}
In this later context the electron spin is not meant to be a key ingredient, contrary to the topological insulators known to date where the strong spin-orbit coupling is responsible for the non trivial topology. Focusing in the two-dimensional (2D) case, and using spinless fermionic models, the strategy then is to search for spontaneously broken $\mathcal{T}$ phases showing an AH or QAH effect driven by Coulomb interactions.

A realization of the Haldane model through electron-electron interactions was obtained at mean field level in the honeycomb lattice  in refs. \onlinecite{RQetal08,WF10}  by adding second neighbor Coulomb interactions. Other proposals involve more complex lattices which allow for intracell fluxes, as the checkerboard, the Kagome, or the decorated honeycomb.\cite{SYetal09,LYM10,WRetal10} $\mathcal{T}$--broken superconducting states have also been proposed recently on the honeycomb lattice.\cite{Li11,NLC11} 

In a recent publication \cite{CGetal11}
we proposed enlarging the unit cell of simple lattice models as 
an alternative way to drive the spontaneous appearance of phases with broken
$\mathcal{T}$.  We explored as an example the nearest neighbor (NN) tight binding model for spinless fermions interacting through a NN Coulomb interaction in the honeycomb lattice. Our main motivation was the idea that enlarging the unit cell -- to enable for instance Kekul\'e type of distortions -- would allow for non--trivial topological phases without the necessity to go to longer range hopping or interactions. We also focussed on a region of high doping near the Van Hove singularity where short range electronic interactions are enhanced and give rise to interesting phases.
For this purpose we considered a minimal model and restricted the mean field decoupling to order parameters of the type $<a^+_i b_j>$ ignoring possible charge ordered phases with order parameters of the type $<c^+_i c_i>$. The result was that $\mathcal{T}$--broken phases with interesting topological features appeared at large fillings above the Van Hove filling for reasonable values of the interaction. The most interesting region of the phase diagram occurred around  the commensurate value $n=1$ corresponding to have one electrons per enlarged (six atoms) unit cell. There the system was insulating above a critical value of the NN Hubbard interaction $V\equiv V_1$. 
In this work we complete the former analysis by allowing charge decouplings in the mean field equations. We find that charge inhomogeneous phases dominate the phase diagram in the region where $\mathcal{T}$ broken phases occurred. $\mathcal{T}$ broken  phases reemerge and are stabilized by the addition of a next nearest neighbor (NNN)interaction $V_2$. The results then are similar to these obtained at half filling in the pioneer work of of Ref.~\onlinecite{RQetal08} in the sense that $V_2 \neq 0$ is needed to stabilize non-trivial phases. In the present work, however, we have found that the high doping $\mathcal{T}$--broken phases are stabilized for $V_2 \ll V_1$ while at half filling $V_2 > V_1$ is required.


The article is organized as follows:  In section \ref{sec_model} we describe the model and the method of calculation. In Sec.  \ref{sec_PD} we explain the phase diagram where for completeness, and to compare with previous results, we present also the situation at half filling (\ref{half}). In Sec. \ref{hf} we  analyze the  modification introduced 
in the $V_{2}=0$ case by the
charge decoupling at higher fillings. We will see that the charge modulated phases wash out
the topologically non--trivial phases. Finally we  see how these are restored by the
inclusion of the second neighbor interaction and  describe the
$\mathcal{T}$ broken  phases and their band structure. 
In Sec. \ref{conclusion} we summarize the situation and discuss some open problems. The Appendix \ref{MFA} contains the
technical details of the calculation.

\section{Model}
\label{sec_model}
\begin{figure}
\begin{centering}
\includegraphics[width=0.9\columnwidth]{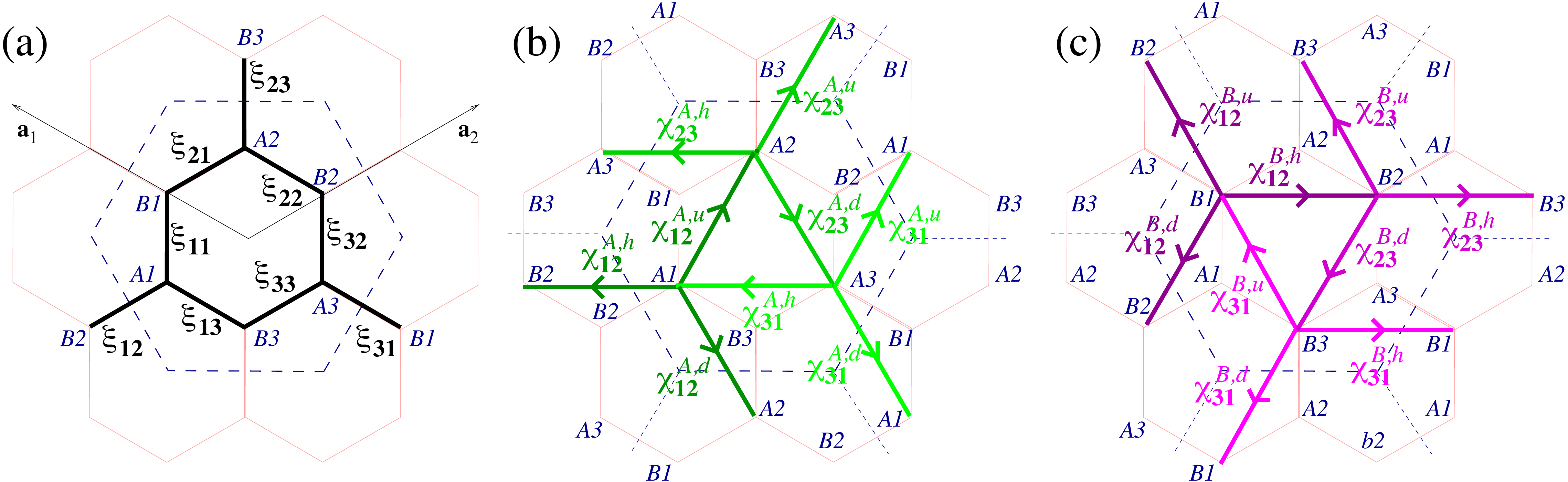}
\par\end{centering}
\caption{\label{fig:uc}Unit cell and mean field parameters of our model. In each
panel we show 9 distinct effective hoppings, making up a total of
27 (see appendix \ref{MF}).}
\end{figure}

We consider the model describing spinless electrons on the honeycomb
lattice,
\begin{equation}
H=-t\sum_{\left\langle i,j\right\rangle }c_{i}^{\dagger}c_{j}+
V_{1}\sum_{\left\langle i,j\right\rangle }n_{i}n_{j}+
V_{2}\sum_{\left\langle \left\langle i,j\right\rangle \right\rangle }n_{i}n_{j}\,
\label{eq:H}
\end{equation}
where $t$ is the NN hopping, and $V_{1}$ and $V_{2}$ the NN and
NNN repulsion, all of them real. The operator
$c_{i}^{\dagger}$ ($c_{i}$) creates (annihilates) a fermion at site
$i$, the number operator is $n_{i}=c_{i}^{\dagger}c_{i}$, and the
sums run over either NN sites $\left\langle i,j\right\rangle $ or
NNN sites $\left\langle \left\langle i,j\right\rangle \right\rangle $.


We use a $6-$atom unit cell to allow for finite flux also in NN loops,
and not only in NNN loops. The basis vectors can be chosen as $\mathbf{a}_{1}=\frac{3a}{2}(-\sqrt{3},1)$
and $\mathbf{a}_{2}=\frac{3a}{2}(\sqrt{3},1)$, and their counterparts
in reciprocal space are $\mathbf{b}_{1}=\frac{2\pi}{3\sqrt{3}a}(-1,\sqrt{3})$
and $\mathbf{b}_{2}=\frac{2\pi}{3\sqrt{3}a}(1,\sqrt{3})$. The number
of Fourier components of local operators is six, $a_{\iota,\mathbf{k}}^{\dagger}=\frac{1}{\sqrt{N}}\sum_{i\in A\iota}c_{i}^{\dagger}e^{i\mathbf{k}.\cdot\mathbf{r}_{i}}$
and $b_{\iota,\mathbf{k}}^{\dagger}=\frac{1}{\sqrt{N}}\sum_{i\in B\iota}c_{i}^{\dagger}e^{i\mathbf{k}.\cdot\mathbf{r}_{i}}$,
with $\iota=1,2,3$. In terms of the new operators the Hamiltonian
in Eq.~\eqref{eq:H} reads as\begin{align}
H & =-t\sum_{\mathbf{k}}a_{i,\mathbf{k}}^{\dagger}(\gamma_{\mathbf{k}}^{ij})^{*}b_{j,\mathbf{k}}+\mbox{H.c.}\nonumber \\
 & +\frac{V_{1}}{N}\sum_{\mathbf{k},\mathbf{k}',\mathbf{q}}a_{i,\mathbf{k}}^{\dagger}a_{i,\mathbf{k}-\mathbf{q}}\gamma_{\mathbf{q}}^{ij}b_{j,\mathbf{k}'}^{\dagger}b_{j,\mathbf{k}'+\mathbf{q}}\nonumber \\
 & +\frac{V_{2}}{2N}\sum_{\mathbf{k},\mathbf{k}',\mathbf{q}}a_{i,\mathbf{k}}^{\dagger}a_{i,\mathbf{k}-\mathbf{q}}\alpha_{\mathbf{q}}^{ij}a_{j,\mathbf{k}'}^{\dagger}a_{j,\mathbf{k}'+\mathbf{q}}\nonumber \\
 & +\frac{V_{2}}{2N}\sum_{\mathbf{k},\mathbf{k}',\mathbf{q}}b_{i,\mathbf{k}}^{\dagger}b_{i,\mathbf{k}-\mathbf{q}}\beta_{\mathbf{q}}^{ij}b_{j,\mathbf{k}'}^{\dagger}b_{j,\mathbf{k}'+\mathbf{q}}\,,\label{eq:H6atom}\end{align}
where summation over repeated indices is assumed (thus the factor
$1/2$ in the last two terms), with $i,j=1,2,3$, and where we have
defined the $3\times3$ matrices

\begin{eqnarray}
\bm{\gamma}_{\mathbf{q}} & = & \left[\begin{array}{ccc}
1 & e^{-i\mathbf{a}_{2}\cdot\mathbf{q}} & 1\\
1 & 1 & e^{i(\mathbf{a}_{1}+\mathbf{a}_{2})\cdot\mathbf{q}}\\
e^{-i\mathbf{a}_{1}\cdot\mathbf{q}} & 1 & 1\end{array}\right]\,,\label{eq:gamma3x3}\\
\bm{\alpha}_{\mathbf{q}} & = & \left[\begin{array}{ccc}
0 & 1+e^{i\mathbf{q}\cdot(\mathbf{a}_{1}+\mathbf{a}_{2})}+e^{i\mathbf{q}\cdot\mathbf{a}_{2}} & 1+e^{-i\mathbf{q}\cdot\mathbf{a}_{1}}+e^{i\mathbf{q}\cdot\mathbf{a}_{2}}\\
1+e^{-i\mathbf{q}\cdot(\mathbf{a}_{1}+\mathbf{a}_{2})}+e^{-i\mathbf{q}\cdot\mathbf{a}_{2}} & 0 & 1+e^{-i\mathbf{q}\cdot(\mathbf{a}_{1}+\mathbf{a}_{2})}+e^{-i\mathbf{q}\cdot\mathbf{a}_{1}}\\
1+e^{i\mathbf{q}\cdot\mathbf{a}_{1}}+e^{-i\mathbf{q}\cdot\mathbf{a}_{2}} & 1+e^{i\mathbf{q}\cdot(\mathbf{a}_{1}+\mathbf{a}_{2})}+e^{i\mathbf{q}\cdot\mathbf{a}_{1}} & 0\end{array}\right]\,,\label{eq:alpha3x3}\\
\bm{\beta}_{\mathbf{q}} & = & \left[\begin{array}{ccc}
0 & 1+e^{-i\mathbf{q}\cdot\mathbf{a}_{1}}+e^{i\mathbf{q}\cdot\mathbf{a}_{2}} & 1+e^{-i\mathbf{q}\cdot(\mathbf{a}_{1}+\mathbf{a}_{2})}+e^{-i\mathbf{q}\cdot\mathbf{a}_{1}}\\
1+e^{i\mathbf{q}\cdot\mathbf{a}_{1}}+e^{-i\mathbf{q}\cdot\mathbf{a}_{2}} & 0 & 1+e^{-i\mathbf{q}\cdot(\mathbf{a}_{1}+\mathbf{a}_{2})}+e^{-i\mathbf{q}\cdot\mathbf{a}_{2}}\\
1+e^{i\mathbf{q}\cdot(\mathbf{a}_{1}+\mathbf{a}_{2})}+e^{i\mathbf{q}\cdot\mathbf{a}_{1}} & 1+e^{i\mathbf{q}\cdot(\mathbf{a}_{1}+\mathbf{a}_{2})}+e^{i\mathbf{q}\cdot\mathbf{a}_{2}} & 0\end{array}\right]\,.\label{eq:beta3x3}\end{eqnarray}
Note that while $\bm{\alpha}_{\mathbf{q}}$ and $\bm{\beta}_{\mathbf{q}}$
are hermitian, the matrix $\bm{\gamma}_{\mathbf{q}}$ is not.


\section{Phase diagram}
\label{sec_PD}
The phase diagram of the model is obtained within the variational mean field approach. The details of the mean field decoupling and related equations are extensively explained in Appendix~A. In brief, we replace the four fermion interaction terms in Eq.~(2) with bilinears which, when written in real space, can be interpreted as the most general hopping and potential energy terms compatible with the reduced translational symmetry of the lattice (6 atom unit cell). In total there are 9 NN ($\xi$) and 18 NNN (9 per sublattice, $\chi^A$ and $\chi^B$) complex hopping parameters, which are depicted in Fig.~1. In addition there are 6 local energy terms (3 per sublattice, $\rho^A$ and $\rho^B$), of which only 5 are independent due to charge conservation. See Appendix~A1 for details.

Using the variational mean field approach one finds the set of $33=3\times 9+6$ mean field equations which, complemented by charge conservation, determine the mean field parameters and the chemical potential $\mu$ (see Appendix~A2 and~A3). The mean field equations read as,
\begin{eqnarray}
\xi_{ij}&=&-\frac{V_1}{N}\sum_{\mathbf{k}}\gamma_{\mathbf{k}}^{ij}
\langle b_{j}^{\dagger}(\mathbf{k})a_{i}(\mathbf{k})\rangle _{MF},\\
\chi^{A,\delta}_{ij}&=&-\frac{V_2}{N}\sum_{\mathbf{k}}\lambda^{A,\delta}_{\mathbf{k},ij}
\langle a_{j}^{\dagger}(\mathbf{k})a_{i}(\mathbf{k})\rangle _{MF},\\
\chi^{B,\delta}_{ij}&=&-\frac{V_2}{N}\sum_{\mathbf{k}}\lambda^{B,\delta}_{\mathbf{k},ij}
\langle b_{j}^{\dagger}(\mathbf{k})b_{i}(\mathbf{k})\rangle _{MF},\\
\rho^{A}_{i}&=&V_1 n_{B}+3V_2 n_{A}-3V:2 n^{A}_{i},\\
\rho^{B}_{i}&=&V_1 n_{A}+3V_2 n_{B}-3V_2 n^{B}_{i},
\label{eq:MFVVpCDW}
\end{eqnarray}
where $\lambda^{A,\delta}_{\mathbf{k},ij}$, $\lambda^{B,\delta}_{\mathbf{k},ij}$ are 
phase factors  analogous to $\gamma_{k}^{ij}$ defined in Eq~(3), $n^{c}_{i}=\frac{1}{N}\sum_{k}\langle c_{i}^{\dagger}(\mathbf{k})c_{i}(\mathbf{k})\rangle _{MF}$ and $n_{c}=\sum^{3}_{i=1}n^{c}_{i}$ with $c=A,B$. Detailed expressions for these matrices can be found in  the appendix \ref{MFE}. 
The notation $\left\langle \dots\right\rangle _{MF}$ means average in the macrocanonical ensamble taking the mean field Hamiltonian in the Boltzmann factor.\\

In order to obtain the mean field phase diagram we solve the mean field equations self consistently, (see Appendix A3) 
and take the solution (if more than one is obtained) which minimizes the free energy in Eq.~\eqref{eq:feVar} (see Appendix A4). Care must be taken with charge like order parameters, Eqs. (9) and (10). Due to the frustration introduced by NNN interaction, these order parameters may flow to a non-selfconsistent solution where the charge like order parameters in different sublattices interchange at each step. Apart from this subtlety, getting a solution is straightforward.

We will analyze first the phase diagram obtained at half filling which
is interesting on its own and later discuss the  modification introduced 
in the $V_{2}=0$ case by the
charge decoupling. We will see that the charge modulated phases wash out
the topologically non--trivial phases. Finally we  see how these are restored by the
inclusion of the
second neighbor interaction.

\subsection{Half-filling}
\label{half}
\begin{figure}

\begin{centering}
\includegraphics[width=0.42\columnwidth]{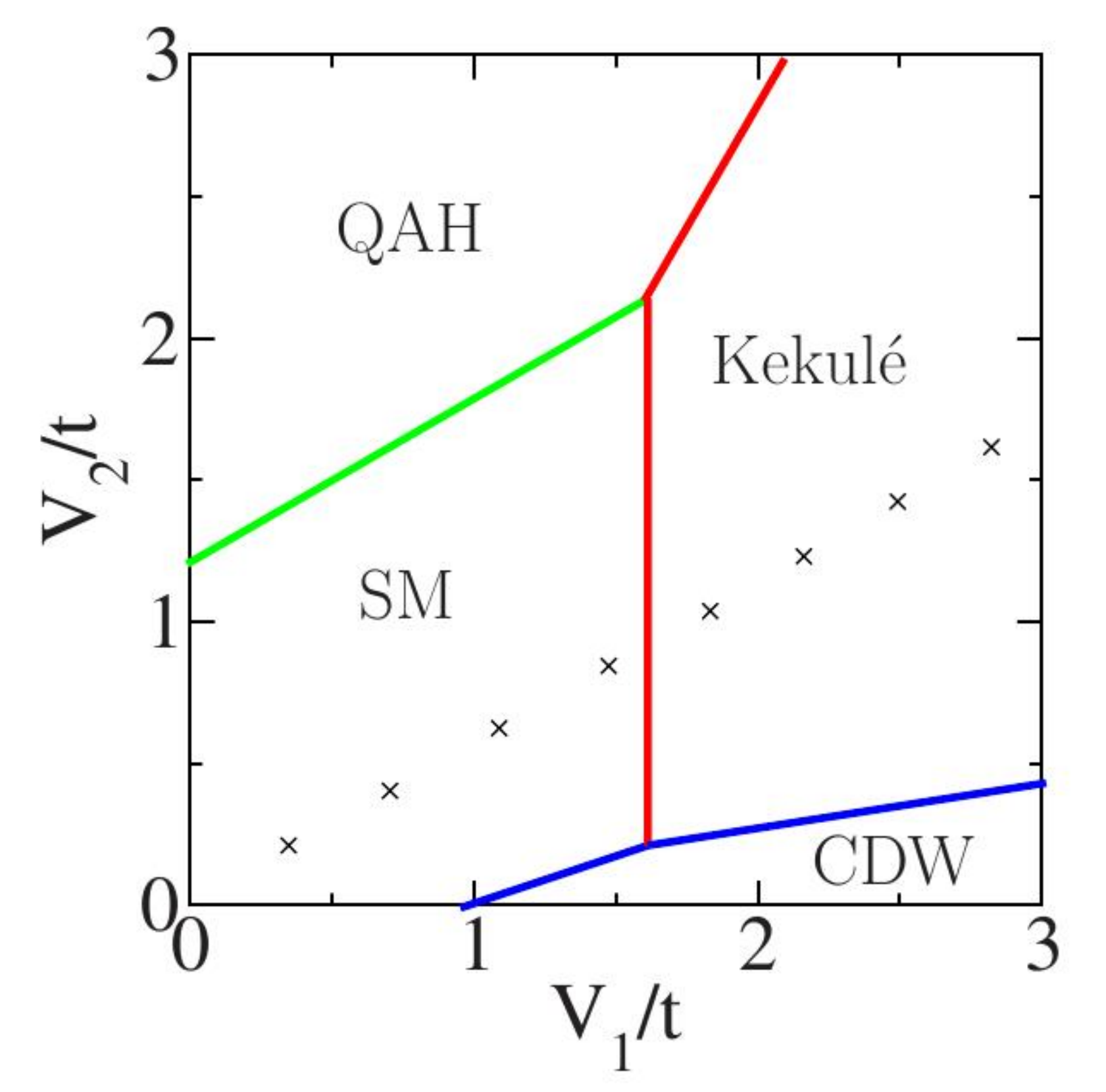}~~\includegraphics[width=0.55\columnwidth]{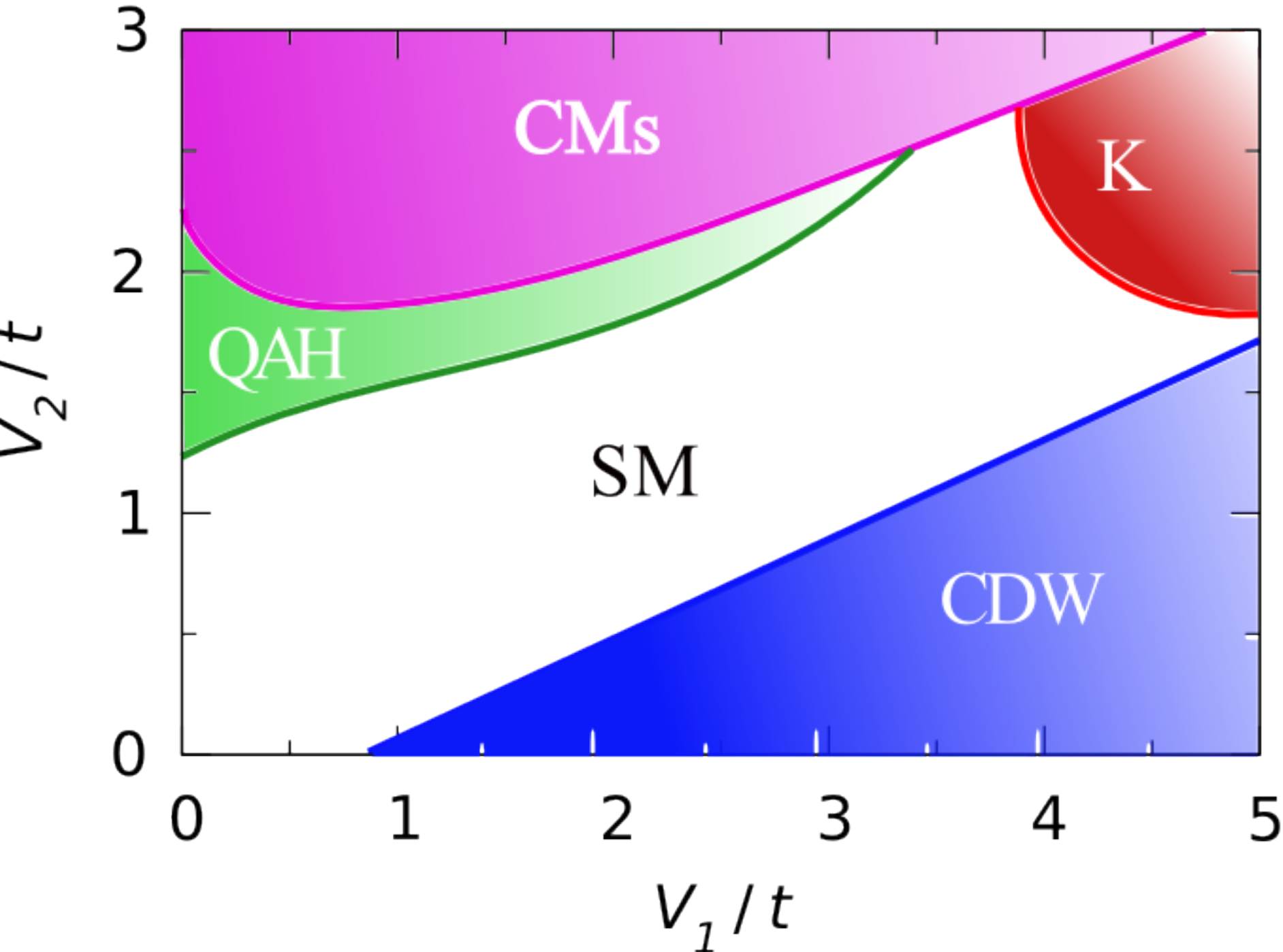}
\par\end{centering}

\caption{\label{fig:pdHalfFil}(Left) Mean field phase diagram for the half filling case reproduced from
ref.~\onlinecite{WF10}. The various phases are described in the text. SM means semi-metal. (Right) Mean field phase diagram obtained in
present work. Lines are guides to the eyes.  CMs stands for the charge modulated phase discussed in the text.}

\end{figure}

Let us first analyze the half-filled case, where $n\equiv n_A + n_B - 3 =0$. This case
provides a test to the present mean field analysis, since a similar
approach, also using a $6-$atom unit cell, has been taken in ref.~\onlinecite{WF10}.
For comparison, we show the phase diagram obtained in ref.~\onlinecite{WF10}  in
the left panel of Fig.~\ref{fig:pdHalfFil}. In the right panel of
Fig.~\ref{fig:pdHalfFil} we can see the phase diagram of the present
work (we use the same color code). We plot the different phases (that will be described in what follows) 
as a function of the interaction strength $V_1$ and $V_2$ in units of the hopping parameter $t$.  
The half filled case was first explored in the original lattice
in ref. \onlinecite{RQetal08} and non--trivial topological phases were already encountered for values of
the interaction $V_2>V_1$.

For $V_{1}\lesssim1.5t$ and $V_{2}\lesssim2t$
the two phase diagrams coincide. For $V_{1}\gtrsim1.5t$, however,
we find that the semi-metallic (SM) and the charge density wave (CDW)
phases are robust against the Kekulé phase. 
The Kekulé phase is characterized by an alternating bond strength as shown schematically
in the inset of the left hand side of Fig. \ref{fig:pdT}.
This distortion is important in the physics of graphene because it opens a gap in the spectrum 
breaking the translational symmetry of the original 
Honeycomb lattice while preserving  
time reversal ($\mathcal{T}$) and inversion 
($\mathcal{I}$) symmetries \cite{MGV07}. It also plays a key role in the
models of charge fractionalization in the Honeycomb lattice \cite{HCM07}.
In our approach the Kekulé
phase only appears at much higher $V_{1}$ and $V_{2}$. For $V_{2}\gtrsim2 t$
a new phase sets in, not predicted in ref.~\onlinecite{WF10}. This
is a charge density wave with reduced rotational symmetry; to distinguish
from CDW we denote it as {\it charge modulated} with modulation also over the
sublattice (CMs). In the CDW there is a charge imbalance between sublattices,
but no inhomogeneity over the sublattice: the charge--like order parameters
$(\rho_{1}^{A},\rho_{1}^{B},\rho_{2}^{A},\rho_{2}^{B},\rho_{3}^{A},\rho_{3}^{B})$
take the form $(\rho,-\rho,\rho,-\rho,\rho,-\rho)$. For CMs, however,
the charge is modulated also over the sublattice and the charge like
order parameters take the form $(\rho,-\rho,\rho,-\rho,-\rho-\Delta,\rho+\Delta)$.
We note that for $V_{2}\gtrsim V_1$ such modulation is naturally expected
from a classical point of view (remember that the Hartree contribution
has a classical interpretation): The staggered charge modulation of
CDW minimizes the energy coming from NN repulsion, but it does not
affect the NNN contribution; for $V_{2}\gtrsim V_{1}$ it becomes
energetically favorable to reverse a NN dimmer, paying the corresponding
NN energy $\propto V_{1}$, but reducing the NNN energetic contribution
$\propto V_{2}$. Such inhomogeneous charge modulation over the sublattice
was not allowed in ref.~\onlinecite{WF10}, and that is the reason
why the QAH phase dominates over a larger region of the phase diagram.

\subsection{Higher filling}
\label{hf}
\begin{figure}
\begin{centering}
\includegraphics[width=0.37\columnwidth]{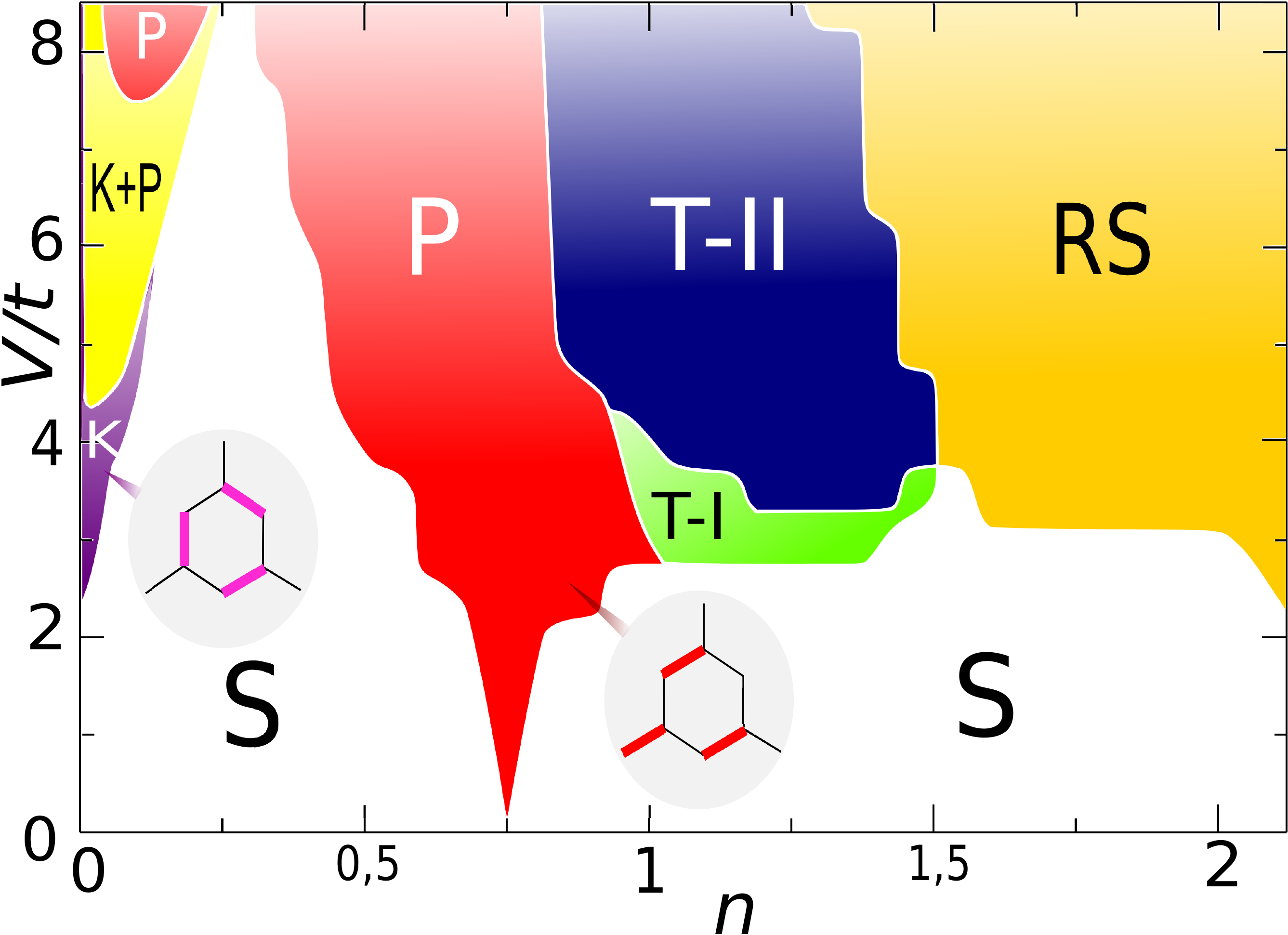} 
\hspace{0.3cm} 
~~\includegraphics[width=0.37\columnwidth]{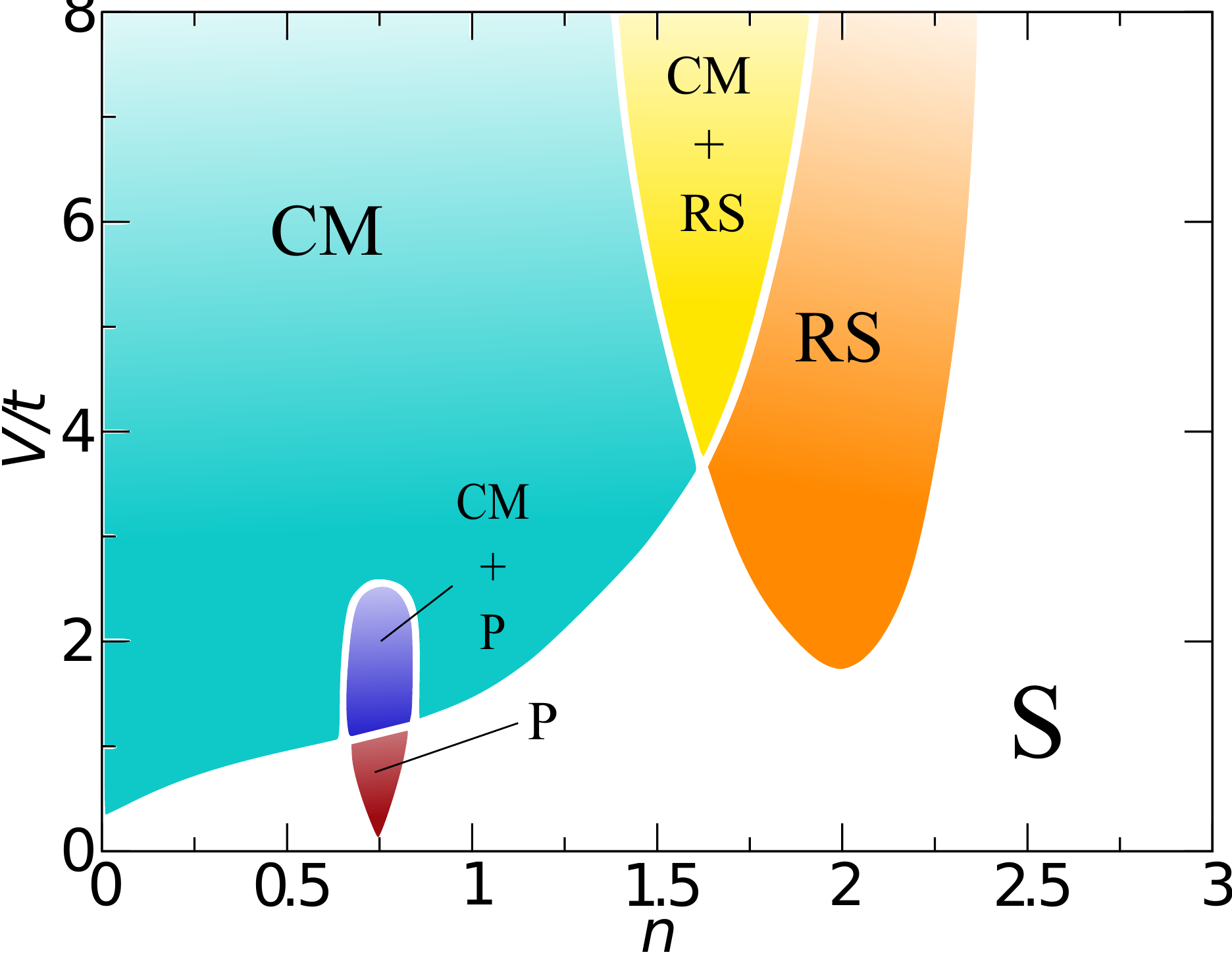}
\par\end{centering}

\caption{\label{fig:pdT}(Left) Phase diagram with $V_{2}=0$ when charge inhomogeneous
phases are not allowed. Legend: (S) symmetric phase,
i.e. bare graphene with a uniform renormalization of the hopping;
(K) Kekul\'e distortion
with hopping renormalization as shown in the inset;
(P) Pomeranchuk distortion of the
Fermi surface and hopping renormalization as shown in the inset;
(K+P) coexistence of Kekule and Pomeranchuk distortions;
(T-I) and (T-II) $\mathcal{T}$ broken  phases discussed at length in the text; 
(RS) broken symmetry state with real hopping parameters, the
distortion is neither Kekul\'e type nor Pomeranchuk (reduced symmetry).
(Right) Phase diagram with $V_{2}=0$ and
charge inhomogeneous phases allowed. CM stands for the new charge modulated phase
discussed in the text. In the CM+RS phase there is also a real, asymmetric renormalization
of the hoppings.
}

\end{figure}

We will now discuss the results obtained for higher dopings. The doped
system has been the subject of attention recently due to the experimental
availability \cite{MBetal10} and to the interesting  phases that emerge
near the Van Hove filling \cite{VV08,MGetal11,NLC11,MSetal11}. 
In ref.~\onlinecite{CGetal11} we have shown that for the Hamiltonian in Eq.~\eqref{eq:H}
with $V_{2}=0$, and ignoring charge inhomogeneous phases, $\mathcal{T}-$broken
phases show up in the phase diagram $V$ vs $n$. 
The phase diagram
is shown in Fig.~\ref{fig:pdT} (left), where the meaning of the various phases are 
described. The dominant phases in the region of
interest above the Van Hove filling  (n=0.75 in our units)  were a Pomeranchuk instability
(P) characterized by rotational symmetry breaking as indicated in the inset, and the 
$\mathcal{T}-$broken
phases denoted by TI and TII obtained at $1\lesssim n\lesssim1.5$ and $V_{1}\gtrsim2t$.
These phases are the same as those obtained  in ref.~\onlinecite{CGetal11}.

When charge inhomogeneous phases are allowed we obtain the phase diagram
shown in Fig.~\ref{fig:pdT} (right). As can be seen, the $\mathcal{T}-$broken
phases are washed out by a charge modulated (CM) phase. This charge
modulation corresponds to a charge imbalance between the two sublattices,
and is homogeneous over the sublattice; at half-filling this is the
CDW discussed in the previous section. This modulation induces a trivial
gap between bands 3 and 4 at the $\Gamma$ point. The system is thus
a trivial insulator at $n=0$, where the Fermi level falls into the
gap, and is metallic for $n>0$. 
We see that the CM phase dominates the region of the phase diagram where Pomeranchuk
and $\mathcal{T}-$broken phases were stabilized before (left hand side of Fig.~\ref{fig:pdT}).
The charge decoupling leaves only a small region around the Van Hove filling (n=0.75) 
and for very small values of the interaction where the Pomeranchuk instability is still the
most favorable phase. At larger values of the interaction (CM+P) the charge modulated
phase is accompanied of a renormalization of the hopping with the Pomeranchuk symmetry. 
This is similar to the  (CM+RS) phase  that occurs in the phase diagram around 
$n=0.75$ for larger values
of the doping and interaction.


A natural way to restore the topologically non--trivial phases is 
to add a NNN interaction $V_2$. 
Since we are looking for topological phases we have centered our attention in
the doping region around the commensurate value $n=1$.

\begin{figure}
\begin{centering}
\includegraphics[width=0.40\columnwidth]{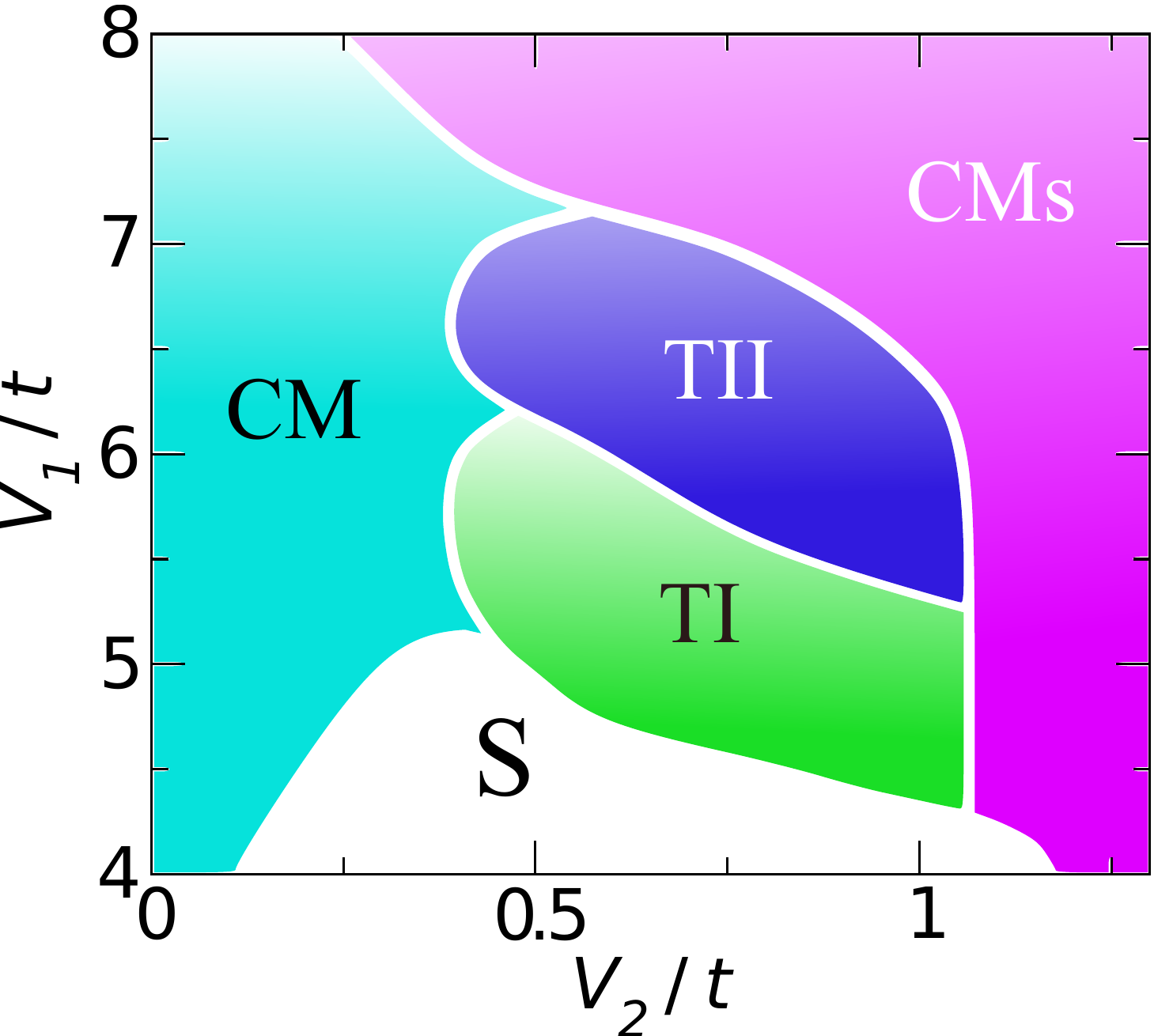}
\par\end{centering}

\caption{\label{fig:v2}Phase diagram in the
plane $V_{1}$ vs $V_{2}$ at $n=1.2$.}

\end{figure}

In Fig.~\ref{fig:v2}  we
show the phase diagram for $V_{2}>0$ obtained at fixed $n=1.2$. 
Increasing $V_{2}$ frustrates the CM phase, and $\mathcal{T}-$broken
phases are recovered. We note that this happens already for $V_{2}\ll V_{1}$.
If $V_{2}$ is further increased the system falls into the CMs phase
(charge modulated with modulation also over the sublattice)
already encountered at half-filling that was discussed in the previous section.

\begin{figure}
\begin{centering}
\includegraphics[width=0.47\columnwidth]{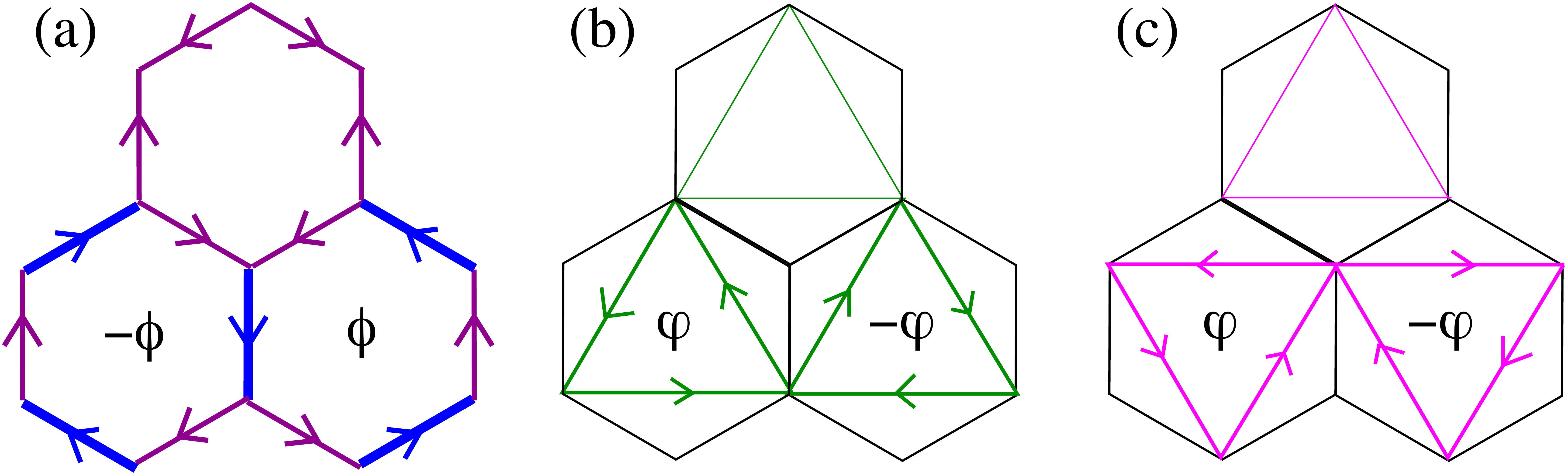}~~\includegraphics[width=0.47\columnwidth]{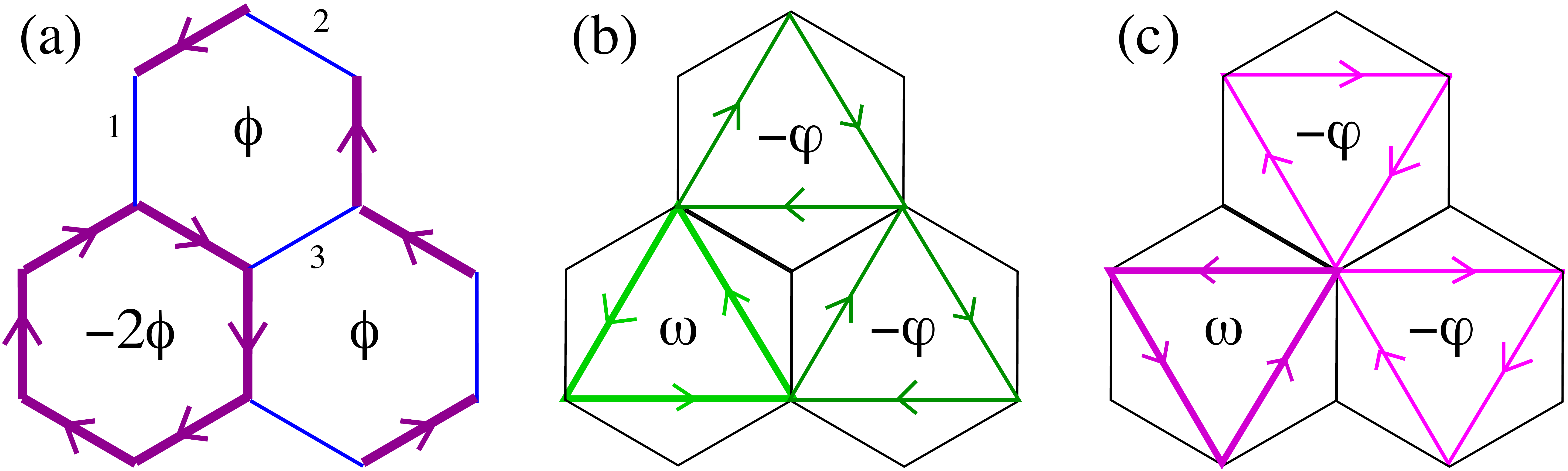}
\par\end{centering}

\caption{\label{fig:fluxesTI}(Left) Flux pattern in the T-I phase. (a): The case with $V_2=0$ and 
no charge decoupling in the phase digram at the left hand side of Fig.~\ref{fig:pdT}. (b) and (c): 
Flux patterns for the T-I phase in Fig.~\ref{fig:v2}. (Right) Same for the T-II phase.}

\end{figure}

The new $\mathcal{T}-$ broken phases T-I
and T-II are similar to the ones described  in ref. \onlinecite{CGetal11}. 
Fig.~\ref{fig:fluxesTI} 
shows the real space hopping pattern for  T-I
and T-II respectively. The direction of the arrows represents the sign of the phase of
the given complex hopping. As discussed in ref. \onlinecite{SF08} the possibility of having  non-trivial topological phases characterized  by a finite Hall conductivity in $\mathcal{T}-$ broken systems is determined by the discrete symmetries $\mathcal{T}-$ and space inversion 
$\mathcal{I}-$ preserved in the system. As happened in the case dicussed in ref.  \onlinecite{CGetal11}, 
the T-II phase is still invariant under inversion hence breaking the product  $\mathcal{T}$.$\mathcal{I}$ and is in principle
topologically non-trivial with finite Hall conductivity \cite{SF08}.

%



\begin{figure}
\begin{centering}
\includegraphics[width=0.29\columnwidth]{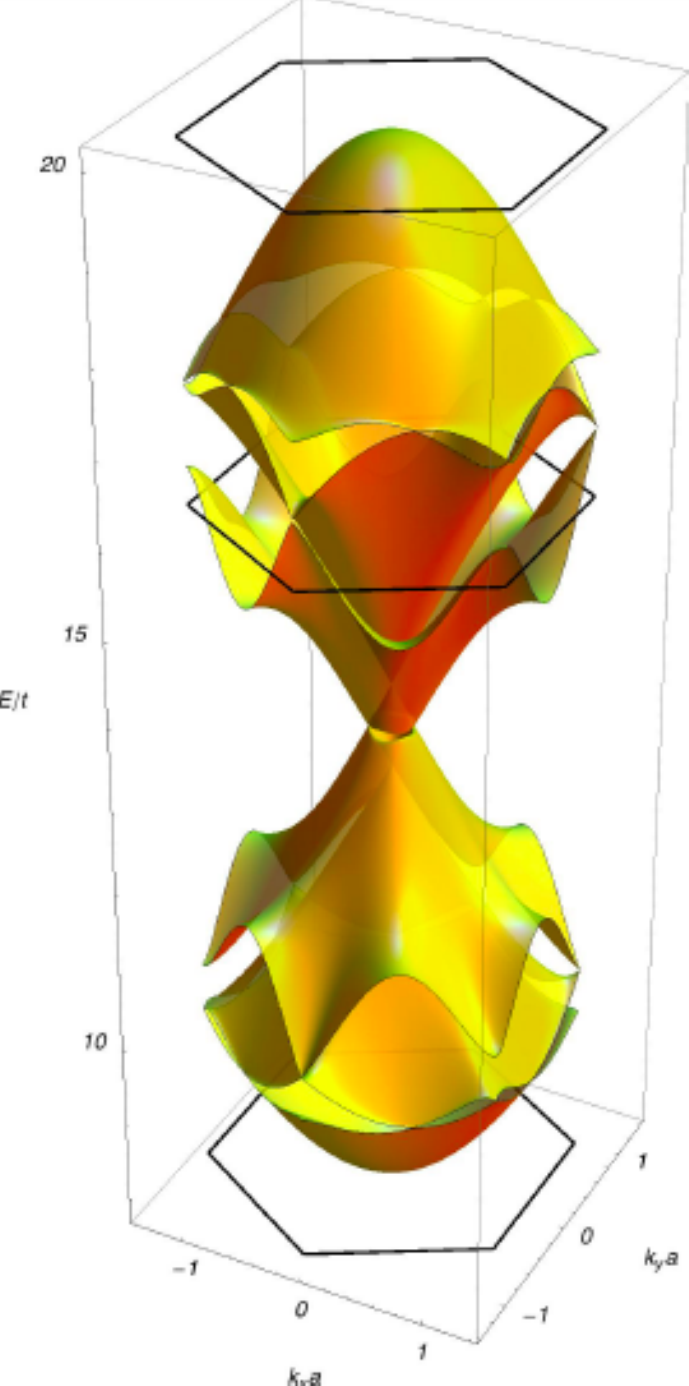}~\includegraphics[width=0.3\columnwidth]{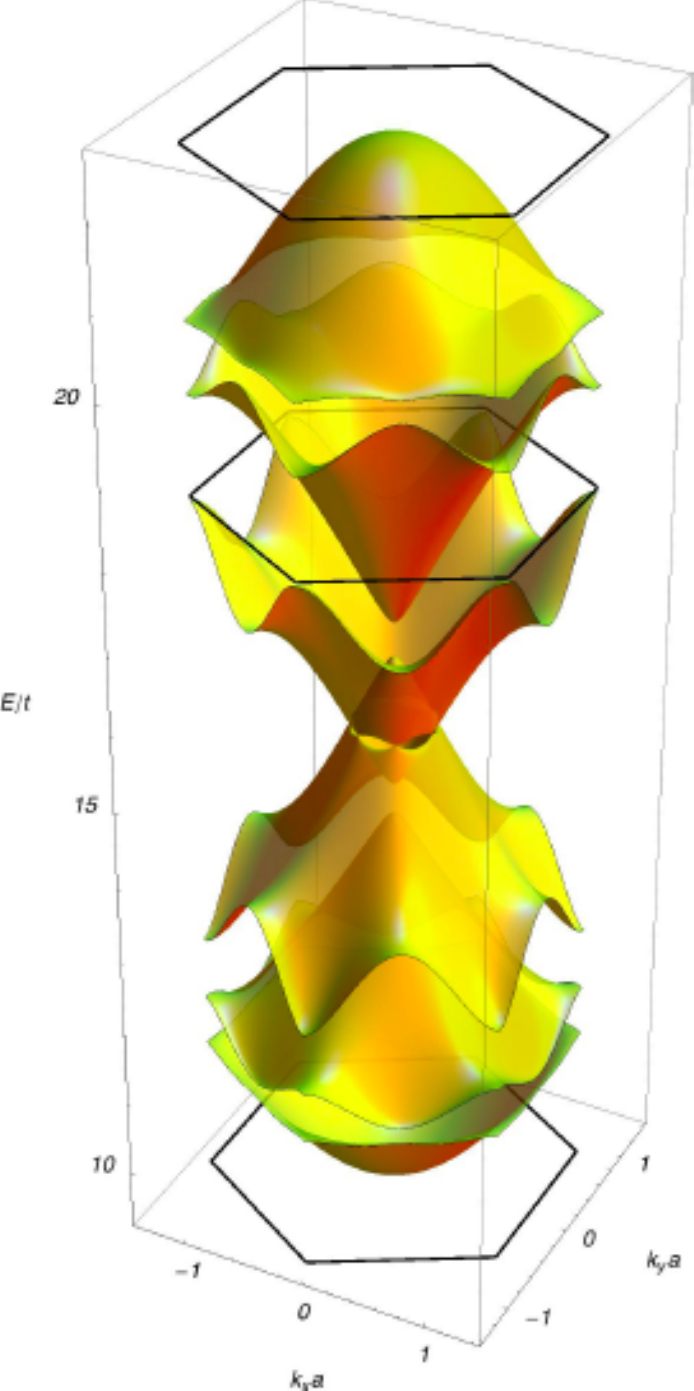}
\par\end{centering}

\caption{\label{fig:bands}Mean field band structure in the T-I phase for $V_{1}=5t$
and $V_{2}=0.75t$ (left) and in the T-II phase for $V_{1}=6t$ and
$V_{2}=0.75t$ (right). The middle black hexagon indicates the position of the Fermi level.}

\end{figure}

Fig.~\ref{fig:bands} shows a  typical band structure for the  T-I 
and T-II phases. The T-I case is shown in Fig.~\ref{fig:bands}(left),
where as an example we took the point $V_{1}=5t$ and $V_{2}=0.75t$.
In the right panel of Fig.~\ref{fig:bands} we show the T-II case,
for $V_{1}=6t$ and $V_{2}=0.75t$. The two figures are very
similar to their counterparts when $V_{2}=0$ \cite{CGetal11}. The interesting feature is that,
unlike what happened there, we do not have any insulating phase away of half filling, even for the
commensurate value $n=1$ whose band structure is very similar to these in  Fig.~\ref{fig:bands}.

\section{Conclusions and discussion}
\label{conclusion}

One of the important points of this and the related work in ref. \onlinecite{CGetal11}  is the  proposal 
that non trivial topological phases can be spontaneously
generated from interactions in lattice fermionic systems when translational
symmetry is relaxed, i.e. when phases with enlarged unit cells are allowed.
The employment of enlarged unit cell
allows for novel  intra-cell current patterns
emerging from interactions, and thus novel topological phases.
We used the Honeycomb lattice with extended Hubbard interaction to exemplify the type of new physics expected. 
In \onlinecite{CGetal11}, we checked \cite{tesis} that no topological phase
can arise in the original lattice (two atoms unit cell), without including a second nearest neighbor interactions $V_2$ to allow for the formation of the necessary current loops that break time reversal symmetry. 

We have completed the analysis of ref. \onlinecite{CGetal11} including a charge decoupling order parameter. The phase diagram is dominated by  charge modulated phases with no particular symmetry and no topological phases were found. 
The charge modulated phases in the case $V_2=0$ are usual CDW phases but we keep the name CM to emphasize the fact that, away from half filling, these phases are metallic. When $V_2\neq 0$ a different charge modulated phase arise that we denote $CMs$ in which
the charge is modulated also over the sublattice and the charge--like
order parameters take the form $(\rho,-\rho,\rho,-\rho,-\rho-\Delta,\rho+\Delta)$. 
An interesting analysis of charge modulated phases in graphene  has been done recently in \cite{GGR12}.

In the previous analysis \cite{CGetal11} there were two special filling values
apart from half filling: The Van Hove filling and the commensurate value $n=1$ corresponding to four electron per enlarged unit cell.
Around the VH density the system supported a Pomeranchuk deformation
and the topological phases were established around the commensurate $n=1$ filling.  The system was gapped only at this particular value. 
In that case we observed a phase separation  in the region 
in the phase space of fillings in between these two values with the two extremes being the stable phases \cite{tesis}. 
We have seeing that in the charge modulated phase dominating the phase diagram the $n=1$ case is not special any more: Even though bands 3 and 4 might be non-trivial now (because of $V_{2}$), the system
at $n=1$ is not  an insulator no ``full'' gap develops as the Fermi
level always crosses bands 4 and 5 (see Fig. \ref{fig:bands}). This behavior can be understood from the strong coupling point of view due to the frustration induced by the competition between $V_1$ and $V_2$ for the ground state. The situation is similar to the quarter filling case in one \cite{baeriswyl03} and two dimensions \cite{fratinimerino09}  where the long-range Coulomb interaction enhances the metallic behavior.

Since this is only a mean field analysis it will be interesting to check the stability of this phases under quantum fluctuations.

Inclusion of an extra coupling to next to nearest neighbors $V_{2}$, restores the topological phases at fillings around $n=1$. This model was already studied in the literature, in particular in ref. \cite{RQetal08} where topological phases were found at half filling for values of the interactions $V_2>V_1$. Allowing the charge decoupling that gives rise to the inhomogeneous charge modulation over the sublattice described along this work, favors the $\mathcal{T}-$ broken phases   that now set at lower values of $V_2$. 
 
The possibility of superconducting phases have been left out in this
analysis.  Considering superconducting order parameters embedded
in enlarged unit cells might make novel topological superconducting phases emerge
\cite{QZ11} opening new routes to realize these, sometimes elusive, phases of matter.
Very recent works suggest the formation of
an eight-atom unit cell spin density wave close to the van Hove filling ( $n\sim 0.75$ in our notation)
\cite{NCC12}, or topological superconducting order exactly at n=0.75 \cite{NLC11}. Whether or not
these phases dominate close to $n\sim 1$ or rather a spin Hall effect is energetically more stable
is still an open question. The $\mathcal{T}-$ broken phases described in this work can be  difficult to observe in graphene since they occur at high values of the interaction and filling
but they can probably be tested in cold atom experiments with optical lattices \cite{W08,SZ08,LGetal09,DMM12}.

\begin{acknowledgments}
This research was supported in part by the Spanish MECD grants FIS2011-23713, FIS2011-29689, PIB2010BZ-00512. F. de J. acknowledges support  from the ``Programa Nacional de Movilidad de Recursos Humanos" (Spanish MECD).
\end{acknowledgments}
\bibliography{Paper3}

\newcommand{\npb}{Nucl. Phys. B}\newcommand{\adv}{Adv.
  Phys.}\newcommand{\epl}{Europhys. Lett.}
\begin{thebibliography}{47}
\expandafter\ifx\csname natexlab\endcsname\relax\def\natexlab#1{#1}\fi
\expandafter\ifx\csname bibnamefont\endcsname\relax
  \def\bibnamefont#1{#1}\fi
\expandafter\ifx\csname bibfnamefont\endcsname\relax
  \def\bibfnamefont#1{#1}\fi
\expandafter\ifx\csname citenamefont\endcsname\relax
  \def\citenamefont#1{#1}\fi
\expandafter\ifx\csname url\endcsname\relax
  \def\url#1{\texttt{#1}}\fi
\expandafter\ifx\csname urlprefix\endcsname\relax\def\urlprefix{URL }\fi
\providecommand{\bibinfo}[2]{#2}
\providecommand{\eprint}[2][]{\url{#2}}

\bibitem[{\citenamefont{Haldane}(2004)}]{H04}
\bibinfo{author}{\bibfnamefont{F.~D.~M.} \bibnamefont{Haldane}},
  \bibinfo{journal}{Phys. Rev. Lett.} \textbf{\bibinfo{volume}{93}},
  \bibinfo{pages}{206602} (\bibinfo{year}{2004}).

\bibitem[{\citenamefont{Haldane}(1988)}]{H88}
\bibinfo{author}{\bibfnamefont{F.~D.~M.} \bibnamefont{Haldane}},
  \bibinfo{journal}{Phys. Rev. Lett.} \textbf{\bibinfo{volume}{61}},
  \bibinfo{pages}{2015} (\bibinfo{year}{1988}).

\bibitem[{\citenamefont{Volovik}(2003)}]{Vo03}
\bibinfo{author}{\bibfnamefont{G.~E.} \bibnamefont{Volovik}},
  \emph{\bibinfo{title}{The universe in a helium droplet}}
  (\bibinfo{publisher}{Clarendon Press, Oxford}, \bibinfo{year}{2003}).

\bibitem[{\citenamefont{Wen}(2004)}]{QFTmbs}
\bibinfo{author}{\bibfnamefont{X.-G.} \bibnamefont{Wen}},
  \emph{\bibinfo{title}{Quantum Field Theory of Many-Body Systems}}
  (\bibinfo{publisher}{Oxford University Press}, \bibinfo{year}{2004}).

\bibitem[{\citenamefont{Bernevig et~al.}(2006)\citenamefont{Bernevig, Hughes,
  and Zhang}}]{BHZ06}
\bibinfo{author}{\bibfnamefont{B.~A.} \bibnamefont{Bernevig}},
  \bibinfo{author}{\bibfnamefont{T.~L.} \bibnamefont{Hughes}},
  \bibnamefont{and} \bibinfo{author}{\bibfnamefont{S.}~\bibnamefont{Zhang}},
  \bibinfo{journal}{Science} \textbf{\bibinfo{volume}{314}},
  \bibinfo{pages}{1757} (\bibinfo{year}{2006}).

\bibitem[{\citenamefont{Hasan and Kane}(2010)}]{HK10}
\bibinfo{author}{\bibfnamefont{M.~Z.} \bibnamefont{Hasan}} \bibnamefont{and}
  \bibinfo{author}{\bibfnamefont{C.~L.} \bibnamefont{Kane}},
  \bibinfo{journal}{Rev. Mod. Phys.} \textbf{\bibinfo{volume}{82}},
  \bibinfo{pages}{3045} (\bibinfo{year}{2010}).

\bibitem[{\citenamefont{Qi and Zhang}(2011)}]{QZ11}
\bibinfo{author}{\bibfnamefont{X.}~\bibnamefont{Qi}} \bibnamefont{and}
  \bibinfo{author}{\bibfnamefont{S.}~\bibnamefont{Zhang}},
  \bibinfo{journal}{Rev. Mod. Phys.} \textbf{\bibinfo{volume}{83}},
  \bibinfo{pages}{1057} (\bibinfo{year}{2011}).

\bibitem[{\citenamefont{Ezawa}(2008)}]{E08}
\bibinfo{author}{\bibfnamefont{Z.~F.} \bibnamefont{Ezawa}},
  \emph{\bibinfo{title}{Quantum Hall Effects - Field Theoretical Approach and
  Related Topics}} (\bibinfo{publisher}{World Scientific, Singapore},
  \bibinfo{year}{2008}).

\bibitem[{\citenamefont{Nagaosa et~al.}(2010)\citenamefont{Nagaosa, Sinova,
  Onoda, MacDonald, and Ong}}]{NSetal10}
\bibinfo{author}{\bibfnamefont{N.}~\bibnamefont{Nagaosa}},
  \bibinfo{author}{\bibfnamefont{J.}~\bibnamefont{Sinova}},
  \bibinfo{author}{\bibfnamefont{S.}~\bibnamefont{Onoda}},
  \bibinfo{author}{\bibfnamefont{A.~H.} \bibnamefont{MacDonald}},
  \bibnamefont{and} \bibinfo{author}{\bibfnamefont{N.~P.} \bibnamefont{Ong}},
  \bibinfo{journal}{Rev. Mod. Phys.} \textbf{\bibinfo{volume}{82}},
  \bibinfo{pages}{1539} (\bibinfo{year}{2010}).

\bibitem[{\citenamefont{Sun and Fradkin}(2008)}]{SF08}
\bibinfo{author}{\bibfnamefont{K.}~\bibnamefont{Sun}} \bibnamefont{and}
  \bibinfo{author}{\bibfnamefont{E.}~\bibnamefont{Fradkin}},
  \bibinfo{journal}{Phys. Rev. B} \textbf{\bibinfo{volume}{78}},
  \bibinfo{pages}{245122} (\bibinfo{year}{2008}).

\bibitem[{\citenamefont{Cortijo et~al.}(2010)\citenamefont{Cortijo, Grushin,
  and Vozmediano}}]{CGV10}
\bibinfo{author}{\bibfnamefont{A.}~\bibnamefont{Cortijo}},
  \bibinfo{author}{\bibfnamefont{A.~G.} \bibnamefont{Grushin}},
  \bibnamefont{and} \bibinfo{author}{\bibfnamefont{M.~A.~H.}
  \bibnamefont{Vozmediano}}, \bibinfo{journal}{Phys. Rev. B}
  \textbf{\bibinfo{volume}{82}}, \bibinfo{pages}{195438}
  (\bibinfo{year}{2010}).

\bibitem[{\citenamefont{Cortijo et~al.}(2012)\citenamefont{Cortijo, Guinea, and
  Vozmediano}}]{CGV12}
\bibinfo{author}{\bibfnamefont{A.}~\bibnamefont{Cortijo}},
  \bibinfo{author}{\bibfnamefont{F.}~\bibnamefont{Guinea}}, \bibnamefont{and}
  \bibinfo{author}{\bibfnamefont{M.~A.~H.} \bibnamefont{Vozmediano}},
  \bibinfo{journal}{J. Phys. A: Math. Theor.} \textbf{\bibinfo{volume}{45}},
  \bibinfo{pages}{383001} (\bibinfo{year}{2012}).

\bibitem[{\citenamefont{Semenoff}(1984)}]{S84}
\bibinfo{author}{\bibfnamefont{G.~V.} \bibnamefont{Semenoff}},
  \bibinfo{journal}{Phys. Rev. Lett.} \textbf{\bibinfo{volume}{53}},
  \bibinfo{pages}{2449} (\bibinfo{year}{1984}).

\bibitem[{\citenamefont{Kane and Mele}(2005{\natexlab{a}})}]{KM05}
\bibinfo{author}{\bibfnamefont{C.}~\bibnamefont{Kane}} \bibnamefont{and}
  \bibinfo{author}{\bibfnamefont{E.}~\bibnamefont{Mele}},
  \bibinfo{journal}{Phys, Rev. Lett.} \textbf{\bibinfo{volume}{95}},
  \bibinfo{pages}{226801} (\bibinfo{year}{2005}{\natexlab{a}}).

\bibitem[{\citenamefont{Kane and Mele}(2005{\natexlab{b}})}]{KM05b}
\bibinfo{author}{\bibfnamefont{C.}~\bibnamefont{Kane}} \bibnamefont{and}
  \bibinfo{author}{\bibfnamefont{E.}~\bibnamefont{Mele}},
  \bibinfo{journal}{Phys, Rev. Lett.} \textbf{\bibinfo{volume}{95}},
  \bibinfo{pages}{146802} (\bibinfo{year}{2005}{\natexlab{b}}).

\bibitem[{\citenamefont{Pesin and Balents}(2010)}]{PB10}
\bibinfo{author}{\bibfnamefont{D.~A.} \bibnamefont{Pesin}} \bibnamefont{and}
  \bibinfo{author}{\bibfnamefont{L.}~\bibnamefont{Balents}},
  \bibinfo{journal}{Nat. Phys.} \textbf{\bibinfo{volume}{6}},
  \bibinfo{pages}{376} (\bibinfo{year}{2010}).

\bibitem[{\citenamefont{Varney et~al.}(2011)\citenamefont{Varney, Sun, Rigol,
  and Galitski}}]{VSetal11}
\bibinfo{author}{\bibfnamefont{C.~N.} \bibnamefont{Varney}},
  \bibinfo{author}{\bibfnamefont{K.}~\bibnamefont{Sun}},
  \bibinfo{author}{\bibfnamefont{M.}~\bibnamefont{Rigol}}, \bibnamefont{and}
  \bibinfo{author}{\bibfnamefont{V.}~\bibnamefont{Galitski}},
  \bibinfo{journal}{Phys. Rev. B} \textbf{\bibinfo{volume}{84}},
  \bibinfo{pages}{241105(R)} (\bibinfo{year}{2011}).

\bibitem[{\citenamefont{Rachel and Hur}(2010)}]{Rachel10}
\bibinfo{author}{\bibfnamefont{S.}~\bibnamefont{Rachel}} \bibnamefont{and}
  \bibinfo{author}{\bibfnamefont{K.~L.} \bibnamefont{Hur}},
  \bibinfo{journal}{Phys. Rev. B} \textbf{\bibinfo{volume}{82}},
  \bibinfo{pages}{075106} (\bibinfo{year}{2010}).

\bibitem[{\citenamefont{Soriano and Fern{\'a}ndez-Rossier}(2010)}]{SR10}
\bibinfo{author}{\bibfnamefont{D.}~\bibnamefont{Soriano}} \bibnamefont{and}
  \bibinfo{author}{\bibfnamefont{J.}~\bibnamefont{Fern{\'a}ndez-Rossier}},
  \bibinfo{journal}{Phys. Rev. B} \textbf{\bibinfo{volume}{82}},
  \bibinfo{pages}{161302} (\bibinfo{year}{2010}).

\bibitem[{\citenamefont{Hohenadler et~al.}(2011)\citenamefont{Hohenadler, Lang,
  and Assaad}}]{HLA11}
\bibinfo{author}{\bibfnamefont{M.}~\bibnamefont{Hohenadler}},
  \bibinfo{author}{\bibfnamefont{T.~C.} \bibnamefont{Lang}}, \bibnamefont{and}
  \bibinfo{author}{\bibfnamefont{F.~F.} \bibnamefont{Assaad}},
  \bibinfo{journal}{Phys. Rev. Lett.} \textbf{\bibinfo{volume}{106}},
  \bibinfo{pages}{100403} (\bibinfo{year}{2011}).

\bibitem[{\citenamefont{Zheng et~al.}()\citenamefont{Zheng, Wu, and
  Zhang}}]{ZWZ10}
\bibinfo{author}{\bibfnamefont{D.}~\bibnamefont{Zheng}},
  \bibinfo{author}{\bibfnamefont{C.}~\bibnamefont{Wu}}, \bibnamefont{and}
  \bibinfo{author}{\bibfnamefont{G.-M.} \bibnamefont{Zhang}},
  \bibinfo{note}{arXiv:1011.5858v2 [cond-mat.str-el]}.

\bibitem[{\citenamefont{Wang et~al.}()\citenamefont{Wang, Shi, Zhang, Wang,
  Dai, and Xie}}]{WSZ+10}
\bibinfo{author}{\bibfnamefont{L.}~\bibnamefont{Wang}},
  \bibinfo{author}{\bibfnamefont{H.}~\bibnamefont{Shi}},
  \bibinfo{author}{\bibfnamefont{S.}~\bibnamefont{Zhang}},
  \bibinfo{author}{\bibfnamefont{X.}~\bibnamefont{Wang}},
  \bibinfo{author}{\bibfnamefont{X.}~\bibnamefont{Dai}}, \bibnamefont{and}
  \bibinfo{author}{\bibfnamefont{X.~C.} \bibnamefont{Xie}},
  \bibinfo{note}{arXiv:1012.5163v1 [cond-mat.str-el]}.

\bibitem[{\citenamefont{Lee}(2011)}]{hailee11}
\bibinfo{author}{\bibfnamefont{D.-H.} \bibnamefont{Lee}},
  \bibinfo{journal}{Phys. Rev. Lett.} \textbf{\bibinfo{volume}{107}},
  \bibinfo{pages}{166806} (\bibinfo{year}{2011}).

\bibitem[{\citenamefont{Araujo et~al.}(2012)\citenamefont{Araujo, Castro, and
  Sacramento}}]{ACS12}
\bibinfo{author}{\bibfnamefont{M.~A.~N.} \bibnamefont{Araujo}},
  \bibinfo{author}{\bibfnamefont{E.~V.} \bibnamefont{Castro}},
  \bibnamefont{and} \bibinfo{author}{\bibfnamefont{P.~D.}
  \bibnamefont{Sacramento}} (\bibinfo{year}{2012}), \eprint{arXiv:1208.1289}.

\bibitem[{\citenamefont{Raghu et~al.}(2008)\citenamefont{Raghu, Qi, Honerkamp,
  and Zhang}}]{RQetal08}
\bibinfo{author}{\bibfnamefont{S.}~\bibnamefont{Raghu}},
  \bibinfo{author}{\bibfnamefont{X.-L.} \bibnamefont{Qi}},
  \bibinfo{author}{\bibfnamefont{C.}~\bibnamefont{Honerkamp}},
  \bibnamefont{and} \bibinfo{author}{\bibfnamefont{S.-C.} \bibnamefont{Zhang}},
  \bibinfo{journal}{Phys. Rev. Lett.} \textbf{\bibinfo{volume}{100}},
  \bibinfo{pages}{156401} (\bibinfo{year}{2008}).

\bibitem[{\citenamefont{Sun et~al.}(2009)\citenamefont{Sun, Yao, Fradkin, and
  Kivelson}}]{SYetal09}
\bibinfo{author}{\bibfnamefont{K.}~\bibnamefont{Sun}},
  \bibinfo{author}{\bibfnamefont{H.}~\bibnamefont{Yao}},
  \bibinfo{author}{\bibfnamefont{E.}~\bibnamefont{Fradkin}}, \bibnamefont{and}
  \bibinfo{author}{\bibfnamefont{S.~A.} \bibnamefont{Kivelson}},
  \bibinfo{journal}{Phys. Rev. Lett.} \textbf{\bibinfo{volume}{103}},
  \bibinfo{pages}{046811} (\bibinfo{year}{2009}).

\bibitem[{\citenamefont{Liu et~al.}(2010)\citenamefont{Liu, Yao, and
  Ma}}]{LYM10}
\bibinfo{author}{\bibfnamefont{Q.}~\bibnamefont{Liu}},
  \bibinfo{author}{\bibfnamefont{H.}~\bibnamefont{Yao}}, \bibnamefont{and}
  \bibinfo{author}{\bibfnamefont{T.}~\bibnamefont{Ma}}, \bibinfo{journal}{Phys.
  Rev. B} \textbf{\bibinfo{volume}{82}}, \bibinfo{pages}{045102}
  (\bibinfo{year}{2010}).

\bibitem[{\citenamefont{Wen et~al.}(2010)\citenamefont{Wen, Ruegg, Wang, and
  Fiete}}]{WRetal10}
\bibinfo{author}{\bibfnamefont{J.}~\bibnamefont{Wen}},
  \bibinfo{author}{\bibfnamefont{A.}~\bibnamefont{Ruegg}},
  \bibinfo{author}{\bibfnamefont{C.-C.~J.} \bibnamefont{Wang}},
  \bibnamefont{and} \bibinfo{author}{\bibfnamefont{G.~A.} \bibnamefont{Fiete}},
  \bibinfo{journal}{Phys. Rev. B} \textbf{\bibinfo{volume}{82}},
  \bibinfo{pages}{075125} (\bibinfo{year}{2010}).

\bibitem[{\citenamefont{Weeks and Franz}(2010)}]{WF10}
\bibinfo{author}{\bibfnamefont{C.}~\bibnamefont{Weeks}} \bibnamefont{and}
  \bibinfo{author}{\bibfnamefont{M.}~\bibnamefont{Franz}},
  \bibinfo{journal}{Phys. Rev. B} \textbf{\bibinfo{volume}{81}},
  \bibinfo{pages}{085105} (\bibinfo{year}{2010}).

\bibitem[{\citenamefont{Li}(2011)}]{Li11}
\bibinfo{author}{\bibfnamefont{T.}~\bibnamefont{Li}},
  \bibinfo{journal}{arXiv:1101.1352}  (\bibinfo{year}{2011}).

\bibitem[{\citenamefont{Nandkishore
  et~al.}(2012{\natexlab{a}})\citenamefont{Nandkishore, Levitov, and
  Chubukov}}]{NLC11}
\bibinfo{author}{\bibfnamefont{R.}~\bibnamefont{Nandkishore}},
  \bibinfo{author}{\bibfnamefont{L.}~\bibnamefont{Levitov}}, \bibnamefont{and}
  \bibinfo{author}{\bibfnamefont{A.}~\bibnamefont{Chubukov}},
  \bibinfo{journal}{Nat. Phys.} \textbf{\bibinfo{volume}{8}},
  \bibinfo{pages}{158} (\bibinfo{year}{2012}{\natexlab{a}}).

\bibitem[{\citenamefont{Castro et~al.}(2011)\citenamefont{Castro, Grushin,
  Valenzuela, Vozmediano, Cortijo, and de~Juan}}]{CGetal11}
\bibinfo{author}{\bibfnamefont{E.~V.} \bibnamefont{Castro}},
  \bibinfo{author}{\bibfnamefont{A.~G.} \bibnamefont{Grushin}},
  \bibinfo{author}{\bibfnamefont{B.}~\bibnamefont{Valenzuela}},
  \bibinfo{author}{\bibfnamefont{M.~A.~H.} \bibnamefont{Vozmediano}},
  \bibinfo{author}{\bibfnamefont{A.}~\bibnamefont{Cortijo}}, \bibnamefont{and}
  \bibinfo{author}{\bibfnamefont{F.}~\bibnamefont{de~Juan}},
  \bibinfo{journal}{Phys. Rev. Lett} \textbf{\bibinfo{volume}{107}},
  \bibinfo{pages}{106402} (\bibinfo{year}{2011}).

\bibitem[{\citenamefont{{J. Ma\~nes} et~al.}(2007)\citenamefont{{J. Ma\~nes},
  Guinea, and Vozmediano}}]{MGV07}
\bibinfo{author}{\bibnamefont{{J. Ma\~nes}}},
  \bibinfo{author}{\bibfnamefont{F.}~\bibnamefont{Guinea}}, \bibnamefont{and}
  \bibinfo{author}{\bibfnamefont{M.~A.~H.} \bibnamefont{Vozmediano}},
  \bibinfo{journal}{Phys. Rev. B} \textbf{\bibinfo{volume}{75}},
  \bibinfo{pages}{155424} (\bibinfo{year}{2007}).

\bibitem[{\citenamefont{Hou et~al.}(2007)\citenamefont{Hou, Chamon, and
  Mudry}}]{HCM07}
\bibinfo{author}{\bibfnamefont{C.-Y.} \bibnamefont{Hou}},
  \bibinfo{author}{\bibfnamefont{C.}~\bibnamefont{Chamon}}, \bibnamefont{and}
  \bibinfo{author}{\bibfnamefont{C.}~\bibnamefont{Mudry}},
  \bibinfo{journal}{Phys. Rev. Lett.} \textbf{\bibinfo{volume}{98}},
  \bibinfo{pages}{186809} (\bibinfo{year}{2007}).

\bibitem[{\citenamefont{McChesney et~al.}(2010)}]{MBetal10}
\bibinfo{author}{\bibfnamefont{J.~L.} \bibnamefont{McChesney}}
  \bibnamefont{et~al.}, \bibinfo{journal}{Phys. Rev. Lett.}
  \textbf{\bibinfo{volume}{104}}, \bibinfo{pages}{136803}
  (\bibinfo{year}{2010}).

\bibitem[{\citenamefont{Valenzuela and Vozmediano}(2008)}]{VV08}
\bibinfo{author}{\bibfnamefont{B.}~\bibnamefont{Valenzuela}} \bibnamefont{and}
  \bibinfo{author}{\bibfnamefont{M.~A.~H.} \bibnamefont{Vozmediano}},
  \bibinfo{journal}{New J. Phys.} \textbf{\bibinfo{volume}{10}},
  \bibinfo{pages}{113009} (\bibinfo{year}{2008}).

\bibitem[{\citenamefont{Makogon et~al.}(2011)\citenamefont{Makogon, van
  Gelderen, Rold\'an, and {Morais Smith}}}]{MGetal11}
\bibinfo{author}{\bibfnamefont{D.}~\bibnamefont{Makogon}},
  \bibinfo{author}{\bibfnamefont{R.}~\bibnamefont{van Gelderen}},
  \bibinfo{author}{\bibfnamefont{R.}~\bibnamefont{Rold\'an}}, \bibnamefont{and}
  \bibinfo{author}{\bibfnamefont{C.}~\bibnamefont{{Morais Smith}}},
  \bibinfo{journal}{Phys. Rev. B} \textbf{\bibinfo{volume}{84}},
  \bibinfo{pages}{125404} (\bibinfo{year}{2011}).

\bibitem[{\citenamefont{Murthy et~al.}(2012)\citenamefont{Murthy, Shimshoni,
  Shankar, and Fertig}}]{MSetal11}
\bibinfo{author}{\bibfnamefont{G.}~\bibnamefont{Murthy}},
  \bibinfo{author}{\bibfnamefont{E.}~\bibnamefont{Shimshoni}},
  \bibinfo{author}{\bibfnamefont{R.}~\bibnamefont{Shankar}}, \bibnamefont{and}
  \bibinfo{author}{\bibfnamefont{H.~A.} \bibnamefont{Fertig}},
  \bibinfo{journal}{Phys. Rev. B} \textbf{\bibinfo{volume}{85}},
  \bibinfo{pages}{073103} (\bibinfo{year}{2012}).

\bibitem[{\citenamefont{Grushin}(2012)}]{tesis}
\bibinfo{author}{\bibfnamefont{A.~G.} \bibnamefont{Grushin}},
  \bibinfo{journal}{Ph. D. Thesis}  (\bibinfo{year}{2012}).

\bibitem[{\citenamefont{Gopalakrishnan
  et~al.}(2012)\citenamefont{Gopalakrishnan, Ghaemi, and Ryu}}]{GGR12}
\bibinfo{author}{\bibfnamefont{S.}~\bibnamefont{Gopalakrishnan}},
  \bibinfo{author}{\bibfnamefont{P.}~\bibnamefont{Ghaemi}}, \bibnamefont{and}
  \bibinfo{author}{\bibfnamefont{S.}~\bibnamefont{Ryu}} (\bibinfo{year}{2012}),
  \eprint{arXiv:1205.0014}.

\bibitem[{\citenamefont{Valenzuela et~al.}(2003)\citenamefont{Valenzuela,
  Fratini, and Baeriswyl}}]{baeriswyl03}
\bibinfo{author}{\bibfnamefont{B.}~\bibnamefont{Valenzuela}},
  \bibinfo{author}{\bibfnamefont{S.}~\bibnamefont{Fratini}}, \bibnamefont{and}
  \bibinfo{author}{\bibfnamefont{D.}~\bibnamefont{Baeriswyl}},
  \bibinfo{journal}{Phys. Rev. B} \textbf{\bibinfo{volume}{68}},
  \bibinfo{pages}{045112} (\bibinfo{year}{2003}),
  \urlprefix\url{http://link.aps.org/doi/10.1103/PhysRevB.68.045112}.

\bibitem[{\citenamefont{Fratini and Merino}(2009)}]{fratinimerino09}
\bibinfo{author}{\bibfnamefont{S.}~\bibnamefont{Fratini}} \bibnamefont{and}
  \bibinfo{author}{\bibfnamefont{J.}~\bibnamefont{Merino}},
  \bibinfo{journal}{Phys. Rev. B} \textbf{\bibinfo{volume}{80}},
  \bibinfo{pages}{165110} (\bibinfo{year}{2009}),
  \urlprefix\url{http://link.aps.org/doi/10.1103/PhysRevB.80.165110}.

\bibitem[{\citenamefont{Nandkishore
  et~al.}(2012{\natexlab{b}})\citenamefont{Nandkishore, Chern, and
  Chubukov}}]{NCC12}
\bibinfo{author}{\bibfnamefont{R.}~\bibnamefont{Nandkishore}},
  \bibinfo{author}{\bibfnamefont{G.}~\bibnamefont{Chern}}, \bibnamefont{and}
  \bibinfo{author}{\bibfnamefont{A.~V.} \bibnamefont{Chubukov}},
  \bibinfo{journal}{Phys. Rev. Lett.} \textbf{\bibinfo{volume}{108}},
  \bibinfo{pages}{227204} (\bibinfo{year}{2012}{\natexlab{b}}).

\bibitem[{\citenamefont{Wu}(2012)}]{W08}
\bibinfo{author}{\bibfnamefont{C.}~\bibnamefont{Wu}}, \bibinfo{journal}{Phys.
  Rev. Lett.} \textbf{\bibinfo{volume}{101}}, \bibinfo{pages}{186807}
  (\bibinfo{year}{2012}).

\bibitem[{\citenamefont{Shao et~al.}(2008)\citenamefont{Shao, Zhu, Sheng, Xing,
  and Wang}}]{SZ08}
\bibinfo{author}{\bibfnamefont{L.~B.} \bibnamefont{Shao}},
  \bibinfo{author}{\bibfnamefont{S.}~\bibnamefont{Zhu}},
  \bibinfo{author}{\bibfnamefont{L.}~\bibnamefont{Sheng}},
  \bibinfo{author}{\bibfnamefont{D.~Y.} \bibnamefont{Xing}}, \bibnamefont{and}
  \bibinfo{author}{\bibfnamefont{Z.~D.} \bibnamefont{Wang}},
  \bibinfo{journal}{Phys. Rev. Lett.} \textbf{\bibinfo{volume}{101}},
  \bibinfo{pages}{246810} (\bibinfo{year}{2008}).

\bibitem[{\citenamefont{Lee et~al.}(2009)\citenamefont{Lee, Gremaud, Han,
  Englert, and Miniatura}}]{LGetal09}
\bibinfo{author}{\bibfnamefont{K.~L.} \bibnamefont{Lee}},
  \bibinfo{author}{\bibfnamefont{B.}~\bibnamefont{Gremaud}},
  \bibinfo{author}{\bibfnamefont{R.}~\bibnamefont{Han}},
  \bibinfo{author}{\bibfnamefont{B.}~\bibnamefont{Englert}}, \bibnamefont{and}
  \bibinfo{author}{\bibfnamefont{C.}~\bibnamefont{Miniatura}},
  \bibinfo{journal}{Phys. Rev. A} \textbf{\bibinfo{volume}{80}},
  \bibinfo{pages}{043411} (\bibinfo{year}{2009}).

\bibitem[{\citenamefont{Dauphin et~al.}(2012)\citenamefont{Dauphin, Muller, and
  Martin-Delgado}}]{DMM12}
\bibinfo{author}{\bibfnamefont{A.}~\bibnamefont{Dauphin}},
  \bibinfo{author}{\bibfnamefont{M.}~\bibnamefont{Muller}}, \bibnamefont{and}
  \bibinfo{author}{\bibfnamefont{M.~A.} \bibnamefont{Martin-Delgado}},
  \bibinfo{journal}{Phys. Rev. A} \textbf{\bibinfo{volume}{86}},
  \bibinfo{pages}{053618} (\bibinfo{year}{2012}).

\end{thebibliography}

\appendix

\section{Mean field analysis}
\label{MFA}

\subsection{Mean field decoupling}
\label{MF}

The mean field Hamiltonian we propose is

\[
H_{MF}=H_{0}+\sum_{\mathbf{k}}\psi_{\mathbf{k}}^{\dagger}\left(\begin{array}{cccccc}
\rho_{1}^{A} & \xi_{11} & \chi_{12}^{A}(\mathbf{k}) & e^{i\mathbf{k}\cdot\mathbf{a}_{2}}\xi_{12} & (\chi_{31}^{A}(\mathbf{k}))^{*} & \xi_{13}\\
\xi_{11}^{*} & \rho_{1}^{B} & \xi_{21}^{*} & \chi_{12}^{B}(\mathbf{k}) & e^{-i\mathbf{k}\cdot\mathbf{a}_{1}}\xi_{31}^{*} & (\chi_{31}^{B}(\mathbf{k}))^{*}\\
(\chi_{12}^{A}(\mathbf{k}))^{*} & \xi_{21} & \rho_{2}^{A} & \xi_{22} & \chi_{23}^{A}(\mathbf{k}) & e^{-i\mathbf{k}\cdot(\mathbf{a}_{1}+\mathbf{a}_{2})}\xi_{23}\\
e^{-i\mathbf{k}\cdot\mathbf{a}_{2}}\xi_{12}^{*} & (\chi_{12}^{B}(\mathbf{k}))^{*} & \xi_{22}^{*} & \rho_{2}^{B} & \xi_{32}^{*} & \chi_{23}^{B}(\mathbf{k})\\
\chi_{31}^{A}(\mathbf{k}) & e^{i\mathbf{k}\cdot\mathbf{a}_{1}}\xi_{31} & (\chi_{23}^{A}(\mathbf{k}))^{*} & \xi_{32} & \rho_{3}^{A} & \xi_{33}\\
\xi_{13}^{*} & \chi_{31}^{B}(\mathbf{k}) & e^{i\mathbf{k}\cdot(\mathbf{a}_{1}+\mathbf{a}_{2})}\xi_{23}^{*} & (\chi_{23}^{B}(\mathbf{k}))^{*} & \xi_{33}^{*} & \rho_{3}^{B}\end{array}\right)\psi_{\mathbf{k}},\]
with \begin{eqnarray}
\chi_{12}^{A}(\mathbf{k}) & = & \chi_{12}^{A,u}+e^{i\mathbf{k}\cdot(\mathbf{a}_{1}+\mathbf{a}_{2})}\chi_{12}^{A,d}+e^{i\mathbf{k}\cdot\mathbf{a}_{2}}\chi_{12}^{A,h}\nonumber \\
\chi_{23}^{A}(\mathbf{k}) & = & e^{-i\mathbf{k}\cdot(\mathbf{a}_{1}+\mathbf{a}_{2})}\chi_{23}^{A,u}+\chi_{23}^{A,d}+e^{-i\mathbf{k}\cdot\mathbf{a}_{1}}\chi_{23}^{A,h}\nonumber \\
\chi_{31}^{A}(\mathbf{k}) & = & e^{-i\mathbf{k}\cdot\mathbf{a}_{2}}\chi_{31}^{A,u}+e^{i\mathbf{k}\cdot\mathbf{a}_{1}}\chi_{31}^{A,d}+\chi_{31}^{A,h}\,,\label{eq:chikA}\end{eqnarray}
and\begin{eqnarray}
\chi_{12}^{B}(\mathbf{k}) & = & \chi_{12}^{B,h}+e^{i\mathbf{k}\cdot\mathbf{a}_{2}}\chi_{12}^{B,d}+e^{-i\mathbf{k}\cdot\mathbf{a}_{1}}\chi_{12}^{B,u}\nonumber \\
\chi_{23}^{B}(\mathbf{k}) & = & e^{-i\mathbf{k}\cdot\mathbf{a}_{2}}\chi_{23}^{B,h}+\chi_{23}^{B,d}+e^{-i\mathbf{k}\cdot(\mathbf{a}_{1}+\mathbf{a}_{2})}\chi_{23}^{B,u}\nonumber \\
\chi_{31}^{B}(\mathbf{k}) & = & e^{i\mathbf{k}\cdot\mathbf{a}_{1}}\chi_{31}^{B,h}+e^{i\mathbf{k}\cdot(\mathbf{a}_{1}+\mathbf{a}_{2})}\chi_{31}^{B,d}+\chi_{31}^{B,u}\,,\label{eq:chikB}\end{eqnarray}
where $H_{0}$ is the bare Hamiltonian, and we use the spinor notation
$\psi_{\mathbf{k}}^{\dagger}=[a_{1}^{\dagger}(\mathbf{k}),b_{1}^{\dagger}(\mathbf{k}),a_{2}^{\dagger}(\mathbf{k}),b_{2}^{\dagger}(\mathbf{k}),a_{3}^{\dagger}(\mathbf{k}),b_{3}^{\dagger}(\mathbf{k})]$.
In Fig.~\ref{fig:uc} we show how the 27 mean field parameters making
up the Fock contribution can be interpreted as NN or NNN hoppings.
The 6 charge like mean field parameters come from the Hartree contribution
due to charge imbalance between the six sites of the unit cell. Due to
charge conservation only 5 of them are independent.

\subsection{Mean field equations}
\label{MFE}

Minimizing the free energy functional\begin{equation}
F[\xi,\chi,\rho]=\left\langle H\right\rangle _{MF}-T\mathcal{S}_{MF}=\mathcal{F}_{MF}+\left\langle H-H_{MF}\right\rangle _{MF}\simeq\Omega_{MF}+\left\langle H-H_{MF}\right\rangle _{MF}+\mu N_{e},\label{eq:fe}\end{equation}
we obtain the following set of mean field equations:\begin{eqnarray*}
\xi_{ij} & = & -\frac{V_{1}}{N}\sum_{\mathbf{q}}\gamma_{\mathbf{q}}^{ij}\left\langle b_{j,\mathbf{q}}^{\dagger}a_{i,\mathbf{q}}\right\rangle _{MF}\\
\chi_{ij,\mathbf{k}}^{A} & = & -\frac{V_{2}}{N}\sum_{\mathbf{q}}\alpha_{\mathbf{k}-\mathbf{q}}^{ij}\left\langle a_{j,\mathbf{q}}^{\dagger}a_{i,\mathbf{q}}\right\rangle _{MF}\\
\chi_{ij,\mathbf{k}}^{B} & = & -\frac{V_{2}}{N}\sum_{\mathbf{q}}\beta_{\mathbf{k}-\mathbf{q}}^{ij}\left\langle b_{j,\mathbf{q}}^{\dagger}b_{i,\mathbf{q}}\right\rangle _{MF}\\
\rho_{i}^{A} & = & V_{1}n^{B}+3V_{2}n^{A}-3V_{2}n_{i}^{A}\\
\rho_{i}^{B} & = & V_{1}n^{A}+3V_{2}n^{B}-3V_{2}n_{i}^{B},\end{eqnarray*}
where we have defined $n_{i}=\frac{1}{N}\sum_{\mathbf{q}}\left\langle c_{i,\mathbf{q}}^{\dagger}c_{i,\mathbf{q}}\right\rangle _{MF}$
and $n=\sum_{i=1}^{3}n_{i}$, with $c=a,b$.

It is easy to show, using Eqs.~\eqref{eq:alpha3x3}--\eqref{eq:beta3x3} and \eqref{eq:chikA}--\eqref{eq:chikB},
that the mean field equation for the effective NNN hoppings can be
cast in the $\mathbf{k}-$independent form\[
\chi_{ij}^{\Gamma,\delta}=-\frac{V_{2}}{N}\sum_{\mathbf{q}}\lambda_{ij,\mathbf{q}}^{\Gamma,\delta}\left\langle c_{j,\mathbf{q}}^{\dagger}c_{i,\mathbf{q}}\right\rangle _{MF},\]
with \[
\lambda_{12,\mathbf{q}}^{A,u}=1\,\,\,\lambda_{12,\mathbf{q}}^{A,d}=e^{-i\mathbf{q}\cdot(\mathbf{a}_{1}+\mathbf{a}_{2})}\,\,\,\lambda_{12,\mathbf{q}}^{A,h}=e^{-i\mathbf{q}\cdot\mathbf{a}_{2}}\]
\[
\lambda_{23,\mathbf{q}}^{A,u}=e^{i\mathbf{q}\cdot(\mathbf{a}_{1}+\mathbf{a}_{2})}\,\,\,\lambda_{23,\mathbf{q}}^{A,d}=1\,\,\,\lambda_{23,\mathbf{q}}^{A,h}=e^{i\mathbf{q}\cdot\mathbf{a}_{1}}\]
\[
\lambda_{31,\mathbf{q}}^{A,u}=e^{i\mathbf{q}\cdot\mathbf{a}_{2}}\,\,\,\lambda_{31,\mathbf{q}}^{A,d}=e^{-i\mathbf{q}\cdot\mathbf{a}_{1}}\,\,\,\lambda_{31,\mathbf{q}}^{A,h}=1\]
and \[
\lambda_{12,\mathbf{q}}^{B,h}=1\,\,\,\lambda_{12,\mathbf{q}}^{B,d}=e^{-i\mathbf{q}\cdot\mathbf{a}_{2}}\,\,\,\lambda_{12,\mathbf{q}}^{B,u}=e^{i\mathbf{q}\cdot\mathbf{a}_{1}}\]
\[
\lambda_{23,\mathbf{q}}^{B,h}=e^{i\mathbf{q}\cdot\mathbf{a}_{2}}\,\,\,\lambda_{23,\mathbf{q}}^{B,d}=1\,\,\,\lambda_{23,\mathbf{q}}^{B,u}=e^{i\mathbf{q}\cdot(\mathbf{a}_{1}+\mathbf{a}_{2})}\]
\[
\lambda_{31,\mathbf{q}}^{B,h}=e^{-i\mathbf{q}\cdot\mathbf{a}_{1}}\,\,\,\lambda_{31,\mathbf{q}}^{B,d}=e^{-i\mathbf{q}\cdot(\mathbf{a}_{1}+\mathbf{a}_{2})}\,\,\,\lambda_{31,\mathbf{q}}^{B,u}=1\,.\]

\subsection{Solving the mean field equation}


In order to solve the set of mean field equations given above we need
to compute the averages of the form $\left\langle b_{j}^{\dagger}(\mathbf{q})a_{i}(\mathbf{q})\right\rangle _{MF}$.
To calculate these averages we introduce the unitary transformation
$\mathcal{U}$ which diagonalizes $\mathcal{H}_{MF}(\mathbf{k})$,\[
\mathcal{U}\mathcal{H}_{MF}(\mathbf{k})\mathcal{U}^{\dagger}=\mbox{diag}[\varepsilon_{1}(\mathbf{k}),\dots,\varepsilon_{6}(\mathbf{k})].\]
The new operators $c_{\alpha}(\mathbf{k})$ are such that \[
\left[\begin{array}{c}
c_{1}(\mathbf{k})\\
c_{2}(\mathbf{k})\\
\vdots\\
c_{5}(\mathbf{k})\\
c_{6}(\mathbf{k})\end{array}\right]=\mathcal{U}\left[\begin{array}{c}
a_{1}(\mathbf{k})\\
b_{1}(\mathbf{k})\\
\vdots\\
a_{3}(\mathbf{k})\\
b_{3}(\mathbf{k})\end{array}\right],\]
and the average $\left\langle b_{j}^{\dagger}(\mathbf{q})a_{i}(\mathbf{q})\right\rangle _{MF}$
and $\left\langle c_{j}^{\dagger}(\mathbf{q})c_{i}(\mathbf{q})\right\rangle _{MF}$
may be written as\begin{eqnarray*}
\left\langle b_{j}^{\dagger}(\mathbf{q})a_{i}(\mathbf{q})\right\rangle _{MF} & = & \left\langle \sum_{\alpha}U_{\alpha,2j}c_{\alpha}^{\dagger}(\mathbf{q})\sum_{\beta}U_{\beta,2i-1}^{*}c_{\beta}(\mathbf{q})\right\rangle _{MF}\\
 & = & \sum_{\alpha=1}^{6}U_{\alpha,2j}U_{\alpha,2i-1}^{*}\left\langle c_{\alpha}^{\dagger}(\mathbf{q})c_{\alpha}(\mathbf{q})\right\rangle _{MF}\\
 & = & \sum_{\alpha=1}^{6}U_{\alpha,2j}U_{\alpha,2i-1}^{*}f[\varepsilon_{\alpha}(\mathbf{q})],\end{eqnarray*}
and\begin{eqnarray*}
\left\langle a_{j}^{\dagger}(\mathbf{q})a_{i}(\mathbf{q})\right\rangle _{MF} & = & \sum_{\alpha=1}^{6}U_{\alpha,2j-1}U_{\alpha,2i-1}^{*}f[\varepsilon_{\alpha}(\mathbf{q})]\\
\left\langle b_{j}^{\dagger}(\mathbf{q})b_{i}(\mathbf{q})\right\rangle _{MF} & = & \sum_{\alpha=1}^{6}U_{\alpha,2j}U_{\alpha,2i}^{*}f[\varepsilon_{\alpha}(\mathbf{q})],\end{eqnarray*}
 and the densities $\left\langle c_{i}^{\dagger}(\mathbf{q})c_{i}(\mathbf{q})\right\rangle _{MF}$
as\begin{eqnarray*}
\left\langle a_{i}^{\dagger}(\mathbf{q})a_{i}(\mathbf{q})\right\rangle _{MF} & = & \sum_{\alpha=1}^{6}\left|U_{\alpha,2i-1}\right|^{2}f[\varepsilon_{\alpha}(\mathbf{q})]\\
\left\langle b_{i}^{\dagger}(\mathbf{q})b_{i}(\mathbf{q})\right\rangle _{MF} & = & \sum_{\alpha=1}^{6}\left|U_{\alpha,2i}\right|^{2}f[\varepsilon_{\alpha}(\mathbf{q})],\end{eqnarray*}
 with\[
f[\varepsilon_{\alpha}(\mathbf{q})]=\frac{1}{\exp\beta[\varepsilon_{\alpha}(\mathbf{q})-\mu]+1}.\]

Then we can write the set of mean field equations as,\begin{eqnarray}
\xi_{ij} & = & -\frac{V_{1}}{N}\sum_{\mathbf{q}}\gamma_{\mathbf{q}}^{ij}\sum_{\alpha=1}^{6}U_{\alpha,2j}U_{\alpha,2i-1}^{*}f[\varepsilon_{\alpha}(\mathbf{q})]\label{eq:SC}\\
\chi_{ij}^{A,\delta} & = & -\frac{V_{2}}{N}\sum_{\mathbf{q}}\lambda_{ij,\mathbf{q}}^{A,\delta}\sum_{\alpha=1}^{6}U_{\alpha,2j-1}U_{\alpha,2i-1}^{*}f[\varepsilon_{\alpha}(\mathbf{q})]\\
\chi_{ij}^{B,\delta} & = & -\frac{V_{2}}{N}\sum_{\mathbf{q}}\lambda_{ij,\mathbf{q}}^{B,\delta}\sum_{\alpha=1}^{6}U_{\alpha,2j}U_{\alpha,2i}^{*}f[\varepsilon_{\alpha}(\mathbf{q})]\\
\rho_{i}^{A} & = & V_{1}n^{B}+3V_{2}n^{A}-\frac{3V_{2}}{N}\sum_{\mathbf{q}}\sum_{\alpha=1}^{6}\left|U_{\alpha,2i-1}\right|^{2}f[\varepsilon_{\alpha}(\mathbf{q})]\label{eq:SCc}\\
\rho_{i}^{B} & = & V_{1}n^{A}+3V_{2}n^{B}-\frac{3V_{2}}{N}\sum_{\mathbf{q}}\sum_{\alpha=1}^{6}\left|U_{\alpha,2i}\right|^{2}f[\varepsilon_{\alpha}(\mathbf{q})],\label{eq:SCend}\end{eqnarray}
with\begin{eqnarray}
n^{A} & = & \frac{1}{N}\sum_{\mathbf{q}}\sum_{i=1}^{3}\sum_{\alpha=1}^{6}\left|U_{\alpha,2i-1}\right|^{2}f[\varepsilon_{\alpha}(\mathbf{q})]\nonumber \\
n^{B} & = & \frac{1}{N}\sum_{\mathbf{q}}\sum_{i=1}^{3}\sum_{\alpha=1}^{6}\left|U_{\alpha,2i}\right|^{2}f[\varepsilon_{\alpha}(\mathbf{q})].\label{eq:nAnB}\end{eqnarray}
This set of equations has to be solved self-consistently with the
constrain imposed by the Luttinger theorem, which reads (ignoring
logarithmic corrections in fermion number $N_{e}$),\begin{eqnarray*}
3+n\equiv\frac{N_{e}}{N}=\frac{1}{N}\frac{\partial\Omega}{\partial\mu}\approx\frac{1}{N}\frac{\partial\Omega_{MF}}{\partial\mu} & = & \frac{\langle\mathcal{N}\rangle_{MF}}{N}\\
 & = & \frac{1}{N}\sum_{\mathbf{q},\alpha}f[\varepsilon_{\alpha}(\mathbf{q})],\end{eqnarray*}
and from which we get $\mu$ self-consistently.

\subsection{Free energy}

To check the stability of possible phases we need the Free energy
defined in Eq.~\eqref{eq:fe}. For a given set of converged order
parameters we have,\begin{equation}
\mathcal{F}=\Omega_{MF}+\mu N_{e}+\langle\mathcal{H}-\mathcal{H}_{MF}\rangle_{MF},\label{eq:feVar}\end{equation}
where the mean field grand canonical potential is given by \[
\Omega_{MF}=-k_{B}T\sum_{\mathbf{q},\alpha}\ln\left\{ 1+e^{-\beta[\varepsilon_{\alpha}(\mathbf{q})-\mu]}\right\} ,\]
and\begin{eqnarray*}
\langle\mathcal{H}-\mathcal{H}_{MF}\rangle_{MF} & = & V_{1}Nn_{A}n_{B}-\frac{V_{1}}{N}\sum_{i,j}|A_{ij}|^{2}-N\sum_{i}(\rho_{i}^{A}n_{i}^{A}+\rho_{i}^{B}n_{i}^{B})-\left(\sum_{i,j}\xi_{ij}A_{ij}+\mbox{c.c.}\right)\\
 & + & \frac{3}{2}V_{2}Nn_{A}n_{A}-\frac{3}{2}V_{2}N\sum_{i=1}^{3}(n_{i}^{A})^{2}+\frac{3}{2}V_{2}Nn_{B}n_{B}-\frac{3}{2}V_{2}N\sum_{i=1}^{3}(n_{i}^{B})^{2}\\
 & - & \frac{V_{2}}{N}\sum_{\Gamma=\{A,B\}}\sum_{\delta=\{u,d,h\}}\left(\left|D_{12}^{\Gamma,\delta}\right|^{2}+\left|D_{23}^{\Gamma,\delta}\right|^{2}+\left|D_{31}^{\Gamma,\delta}\right|^{2}\right)\\
 & - & \sum_{\Gamma=\{A,B\}}\sum_{\delta=\{u,d,h\}}\left(\chi_{12}^{\Gamma,\delta}\left[D_{12}^{\Gamma,\delta}\right]^{*}+\chi_{23}^{\Gamma,\delta}\left[D_{23}^{\Gamma,\delta}\right]^{*}+\left[\chi_{31}^{\Gamma,\delta}\right]^{*}D_{31}^{\Gamma,\delta}+\mbox{c.c.}\right)\end{eqnarray*}
where \begin{eqnarray*}
A_{ij} & = & \sum_{\mathbf{q},\alpha}(\gamma_{\mathbf{q}}^{ij})^{*}U_{\alpha,2i-1}U_{\alpha,2j}^{*}f[\varepsilon_{\alpha}(\mathbf{q})]\\
D_{ij}^{A,\delta} & = & \sum_{\mathbf{q}}\lambda_{ij,\mathbf{q}}^{A,\delta}\sum_{\alpha=1}^{6}U_{\alpha,2j-1}U_{\alpha,2i-1}^{*}f[\varepsilon_{\alpha}(\mathbf{q})]\\
D_{ij}^{B,\delta} & = & \sum_{\mathbf{q}}\lambda_{ij,\mathbf{q}}^{B,\delta}\sum_{\alpha=1}^{6}U_{\alpha,2j}U_{\alpha,2i}^{*}f[\varepsilon_{\alpha}(\mathbf{q})].\end{eqnarray*}

\end{document}